%% file: main.tex
\documentclass[twocolumn]{aastex63}
\usepackage{aas_macros}
\usepackage{color}
\usepackage{soul}
\usepackage{natbib}
\usepackage{hyperref}
\usepackage{amsmath}
\usepackage{xspace}
\usepackage{lineno}
\usepackage{graphicx}
\usepackage{enumitem}
\usepackage{amssymb}
\usepackage{xifthen}
\usepackage{hyperref}
\usepackage[normalem]{ulem}

\hypersetup{    
  colorlinks      = {true},
  linkcolor       = {blue},
  citecolor       = {blue},
  urlcolor        = {blue},
}

\newcommand{\code}[1]{\texttt{#1}\xspace}

\newcommand{\SSSSS}{${S}^5$\xspace}
\newcommand{\gaia}{\textit{Gaia}\xspace}

\newcommand{\unit}[1]{\ensuremath{\mathrm{\,#1}}\xspace}
\newcommand{\feh}{\unit{[Fe/H]}}

\newcommand{\logg}{\ensuremath{\log\,g}\xspace}

\newcommand{\vhel}         {\mbox{$v_{\mathrm{hel}}$}}

\newcommand{\sigmafeh}         {\mbox{$\sigma_{\mathrm{[Fe/H]}}$}}

\newcommand{\kms}{\unit{km\,s^{-1}}}
\newcommand{\masyr}{\unit{mas\,yr^{-1}}}

\newcommand{\Track}{\ensuremath{\mathrm{Track}}}

\newcommand{\revise}[1]{#1}

\shorttitle{ATLAS-Aliqa Uma Stream}
\shortauthors{Li et al.}

\begin{document}

\title{Broken into Pieces: ATLAS and Aliqa Uma as One Single Stream}



\input{authors.tex}

\begin{abstract}
We present the first spectroscopic measurements of the ATLAS and Aliqa Uma streams from the Southern Stellar Stream Spectroscopic Survey (\SSSSS), in combination with the photometric data from the Dark Energy Survey and astrometric data from \gaia. From the coherence of spectroscopic members in radial velocity and proper motion, we find out that these two systems are \revise{extremely likely to be} one stream with discontinuity in morphology and density on the sky (the ``kink" feature). We refer to this entire stream as the ATLAS-Aliqa Uma stream, or the AAU stream. 
We perform a comprehensive exploration of the effect of baryonic substructures and find that only an encounter with the Sagittarius dwarf $\sim 0.5$ Gyr ago can create a feature similar to the observed ``kink". 
In addition, we also identify two gaps in the ATLAS component associated with the broadening in the stream width (the ``broadening" feature). These gaps have likely been created by small mass perturbers, such as dark matter halos, as the AAU stream is the most distant cold stream known with severe variations in both the stream surface density and the stream track on the sky.
With the stream track, stream distance and kinematic information, we determine the orbit of the AAU stream and find that it has been affected by the Large Magellanic Cloud, resulting in a misalignment between the proper motion and stream track.
Together with the Orphan-Chenab Stream, AAU is the second stream pair that has been found to be a single stream separated into two segments by external perturbation.
\end{abstract}


\section{INTRODUCTION}
\label{intro}


\revise{Searching for the lowest-mass dark matter subhalos is a clear way to differentiate between different dark matter models \citep[e.g.,][]{BuckleyPeter2018}.
The currently preferred cosmological model -- Lambda Cold Dark Matter ($\Lambda$CDM) -- predicts the existence of low-mass dark matter halos down to a ``minimum mass'' as small as $3 \times 10^{-6}\, \mathrm{M}_\odot$ \citep{Hofmann:2001bi, Green2004, Diemand2005}. 
Most alternative dark matter models behave like CDM on large scales, but produce different minimum dark matter halo masses.
For example, warm dark matter models with particle masses at a few keV sharply suppresses halos below a mass of $10^8 \mathrm{M}_\odot$\citep{Bullock2017araa}.
Similarly, fuzzy dark matter models with an ultra-light dark matter particle mass of $\sim 10^{-22} {\rm eV}$ have a minimum subhalo mass of $\sim 10^7 \mathrm{M}_\odot$ \citep{hui2017}.}

The lowest-mass dark matter halos are currently found through observations of the lowest \emph{stellar} mass galaxies, which appear to live in $10^{8{-}9}\,\rm M_\odot$ halos \citep[e.g.,][]{Koposov:2009,Jethwa:2018, Kim2018, Newton2018, Nadler2019, Nadler:2020}. This matches theoretical expectations that baryonic effects like supernova feedback and reionization prevent star formation in halos below this scale \citep[e.g.,][]{Bullock2001}.
Thus, one of the possible ways to probe dark matter halos at $\lesssim10^7\,\rm M_\odot$ is to observe the effects of star-free \emph{dark subhalos} on matter with which they interact \citep[e.g.,][]{Johnston:2002}. 
In a smooth gravitational potential, stellar streams formed by tidal disruption of globular clusters \citep[e.g.,][]{Dehnen2004AJ....127.2753D} would stretch into coherent mostly smooth bands on the sky \citep{Kupper:2010}.
However, a dark subhalo impacting the stream disturbs the smooth stream, forming gaps and wiggles \citep[e.g.,][]{2008ApJ...681...40S, Yoon:2011, Carlberg2013, Erkal:2015b}.

Dozens of thin, kinematically cold stellar streams have been discovered in the Milky Way halo \citep{Grillmair:2016,Shipp:2018, Malhan2018, Ibata2019}, and the most prominent ones have already been examined for evidence of density variations.
Indeed, signatures consistent with $10^6 \mathrm{M}_\odot$ dark halo encounters have already been claimed in the Palomar 5 stream \citep[e.g.,][]{Carlberg:2012, Erkal:2017}
and the GD-1 stream \citep[e.g.,][]{Koposov2010,Carlberg2013,Price-Whelan2018, Bonaca2019,deBoer2019}.
From these streams, the inferred abundance of dark matter subhalos down to $\sim10^6\,\rm M_\odot$ is consistent with the CDM predictions \citep[e.g.,][]{Carlberg:2012, Banik:2019}.
However, baryonic structures like giant molecular clouds \citep{Amorisco2016,Banik:2019},
the Milky Way bar \citep{Pearson2017, Erkal:2017}, spiral arms \citep{Banik:2019}
and the disruption of the progenitor \citep{Webb:2019} 
can also produce stream perturbations that mimic the observational signature of dark subhalos.
It is crucial to find more kinematically cold streams with perturbation signatures and better orbital constraints, which will improve our understanding of the baryonic effects on the streams as well as the impact of the smallest dark matter halos.

In this paper, we show that two recently discovered cold stellar streams -- ATLAS and Aliqa Uma, which were previously thought to be unrelated -- are \revise{extremely likely to be} two components of a single system. The discontinuous on-sky morphology is caused by possible perturbations from either baryons or dark matter halos. 

ATLAS was first discovered as a 12\degr\ long cold stellar stream \citep{Koposov:2014} in the first data release (DR1) of the VST ATLAS survey \citep{VST_ATLAS}. The detected length of the stream was mainly limited to the sky coverage of DR1. It was later analyzed by \citet{Bernard:2016} using Pan-STARRS 1 (PS1) data \citep{PS1}, which extended ATLAS to a total length of 28\degr.  With the first three years of data from the Dark Energy Survey \citep[DES;][]{DES:2016}, \citet{Shipp:2018} recovered 22.6\degr\ of the ATLAS stream within the DES footprint, at a heliocentric distance of 22.9 kpc. 

Aliqa Uma was discovered in \citet{Shipp:2018} in DES at a heliocentric distance of 28.8 kpc, residing at the southern end of the ATLAS stream. Despite the close proximity to the ATLAS stream, the difference in distance modulus and orientation on the sky led the authors to conclude that Aliqa Uma was a distinct stream, rather than an extension of ATLAS.

\begin{figure}
    \centering
    \includegraphics[width=0.49\textwidth]{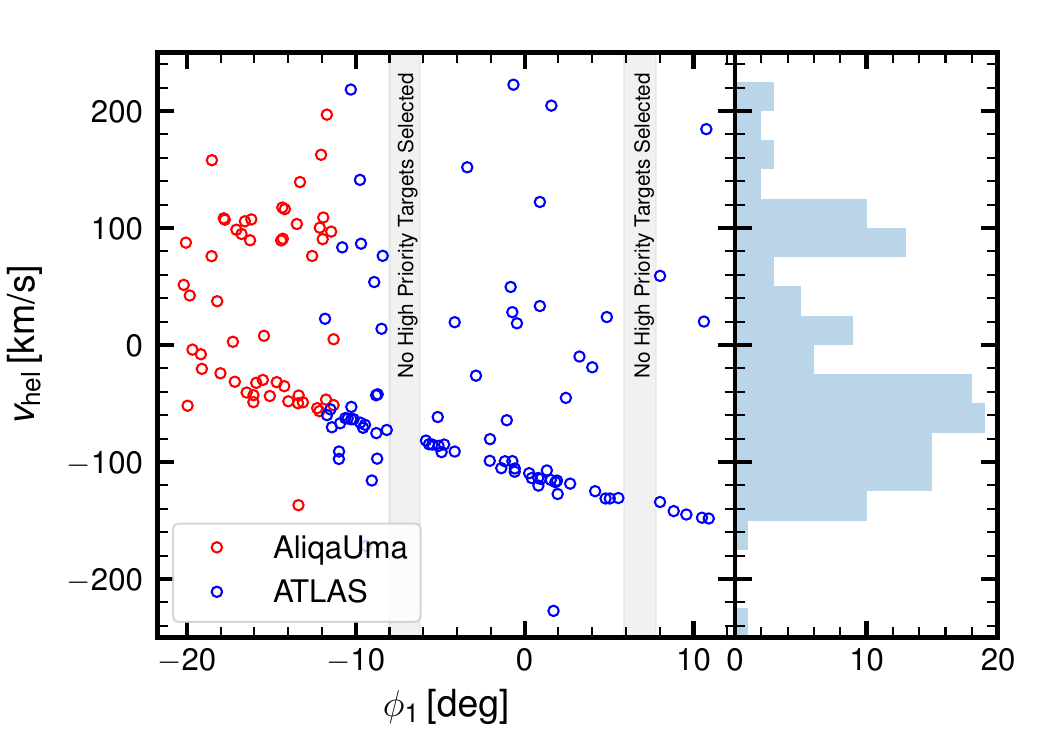}
    \caption{Heliocentric velocity as a function of stream longitude $\phi_1$ for the high-priority targets with $\logg <4.1$ and $\feh < -1$ in the fields of Aliqa Uma (red) and ATLAS (blue) streams. The grey bands show the fields that were observed prior to \SSSSS (Field 8 and 15 in Figure \ref{fig:specsel}) and therefore no high-priority targets were defined. These do not present any true gap in the member star distribution.  The clear spectroscopic members of Aliqa Uma and ATLAS streams follow a coherent velocity trend from $\vhel\sim-20\,\kms$ at $\phi_1\sim-20\degr$ to $\vhel\sim-140\,\kms$ at $\phi_1\sim+10\degr$. We also see additional kinematic substructure in the velocity distribution around $\vhel\sim+100\,\kms$, mostly in the Aliqa Uma stream field (also seen in the histogram in the right panel). We discuss this distinct substructure in Section \ref{sec:palca}.
    }
    \label{fig:vhel_dist}
\end{figure}

Both streams were observed by the Southern Stellar Stream Spectroscopic Survey \citep[\SSSSS;][hereafter Paper I]{Li2019}, which so far has provided 6D phase space information for 12 streams in the Southern Hemisphere with observations taken in 2018 and 2019, by combining AAT/AAOmega spectra with proper motions from \gaia DR2 \citep{GaiaDR2} and photometry from DES DR1 \citep{desdr1}.
As shown in Figure 12 of Paper I and reproduced in Figure \ref{fig:vhel_dist} here, the high priority stream targets (proper motion selected metal-poor candidate members)
in \SSSSS show a clear connection in the line-of-sight velocities for these two streams. Similarly, \citet{Shipp2019} show that the proper motions of the two streams point in nearly the same direction (see Figure 5 in that paper). The kinematic information for the stream members suggests that these two streams \revise{are likely} one stream.  
In this paper, we confirm this hypothesis with kinematics, distances, and metallicities of the stream members, and further explore the physical origins of the discontinuity of the stream track on the sky.

The structure of the paper is as follows. We present the spectroscopic data from \SSSSS in Section \ref{sec:spec}. We then revisit the stream with \gaia DR2 and other deep photometric data including DES DR1 in Section \ref{sec:photometry}. We model the orbital motion of the stream in Section \ref{sec:modeling}. We then discuss different properties of the streams in Section \ref{sec:discuss} and conclude in Section \ref{sec:summary}.

Throughout the paper, we use the rotation matrix for the ATLAS stream defined in \citet{Shipp2019}, also shown in Appendix \ref{sec:coords}, for converting celestial equatorial coordinates ($\alpha, \delta$) to stream coordinates ($\phi_1, \phi_2$). 
We use ($U_\odot$, $W_\odot$) = (11.1, 7.3)~\kms  \citep{Schonrich2010} and $V_\odot=\Omega_\odot R_\odot=245$~\kms \citep{Reid2004, Gravity2019} 
to convert heliocentric velocity ($v_{\mathrm{hel}}$) to velocity in the Galactic standard of rest ($v_{\mathrm{GSR}}$).
Unless otherwise noted, our $gri$ magnitudes are reddening corrected photometry from DES DR1, specifically, taking the \code{WAVG\_MAG\_PSF} quantity corrected with $E(B-V)$ from \citet{Schlegel:1998} and the extinction coefficients from DES DR1.

All paper related materials including data, models and code used in this paper are publically available via GitHub repository.\footnote{\url{https://github.com/s5collab/ATLAS_AliqaUma}}

\section{Spectroscopic Data}\label{sec:spec}

\subsection{\SSSSS Observations}

The ATLAS and Aliqa Uma streams were observed in 2018 as part of the \SSSSS program, which uses the AAOmega spectrograph on the 3.9~m Anglo-Australian Telescope (AAT), fed by the Two Degree Field (``2dF") fiber positioner facility, allowing the acquisition of up to 392 simultaneous spectra of objects within a 2\degr\ field in diameter on the sky. We refer readers to Paper I for details on the survey strategy, target selection, observation and reduction of \SSSSS. We briefly describe the observations and reductions for the two streams here.

A total of 5 AAT fields were observed in Aliqa Uma and 12 fields in ATLAS, with a total covered length of the stream of about 34\degr\ on the sky. Center of each field was separated by $\sim2\degr$. The top panel of Figure~\ref{fig:specsel} shows the 17 AAT fields in ATLAS stream coordinates, denoted as Field 1 to 17. The Aliqa Uma stream is located at $\phi_1 < -9\degr$ (Field 1-5). 
As discussed in Paper I, the track of ATLAS is curved on the sky, and therefore we adopted the polynomial stream track from \citet{Shipp:2018} for the ATLAS stream pointings. Two of the ATLAS fields (Field 8 and 15, encircled in red in the top panel of Figure~\ref{fig:specsel}) were observed prior to \SSSSS as a pilot program, and therefore the target selection strategy, as well as the pointing strategy described in Paper I does not apply to these two fields. In particular, the selection for those two fields was performed without parallax and proper motion information from \gaia DR2. We aligned the rest of ATLAS fields to Field 15, but Field 8 is slightly misaligned, causing a small observational gap in $\phi_1$ coverage around $\phi_1=-8\degr$.

The stream targets are selected using photometry from DES DR1 and astrometry from \gaia DR2. All the targets have been assigned a priority from P9 to P1, with P9 indicating the highest priority. While \SSSSS includes non-stream targets in the observation, stream targets, \revise{which are used to produce Figure \ref{fig:vhel_dist}}, have the highest priority in fiber assignment (P9-P7). The stream targets are selected as either red giant branch stars (RGBs) or blue horizontal branch stars (BHBs) based on their location on the dereddened color-magnitude diagram from DES DR1 photometry. The stream candidates are also selected to have proper motions consistent with measurements in \citet{Shipp2019}. In addition, we put the metal-poor stream member candidates in the highest priority category (P9) based on color-color selection in a $g-r$ vs. $r-i$ diagram \citep[see details in Paper I and ][]{Li2018b}. After all the stream targets are allocated, we use the spare fibers for additional targets in the field, such as RR Lyrae stars, hot stars, extremely metal-poor candidates, and low-redshift galaxy candidates.

The observed data were first reduced and extracted using the \code{2dfdr} pipeline provided by AAO Data Central\footnote{\url{https://www.aao.gov.au/science/software/2dfdr}}. The radial velocity and stellar parameters for each star were then derived by fitting the interpolated synthetic templates from the PHOENIX spectral grid \citep{Husser2013} modified by a polynomial continuum using the \code{RVSpecFit} code \citep{Koposov2011,rvspecfit}.
The means and uncertainties of the radial velocity and stellar parameters are derived from the posterior distribution samples obtained from a Markov Chain Monte Carlo (MCMC) sampler. For stars with multiple observations, the measurements with highest S/N are used.

\subsection{Spectroscopic Member Identification} \label{sec:member}

\begin{figure*}
    \centering
    \includegraphics[width=0.89\textwidth]{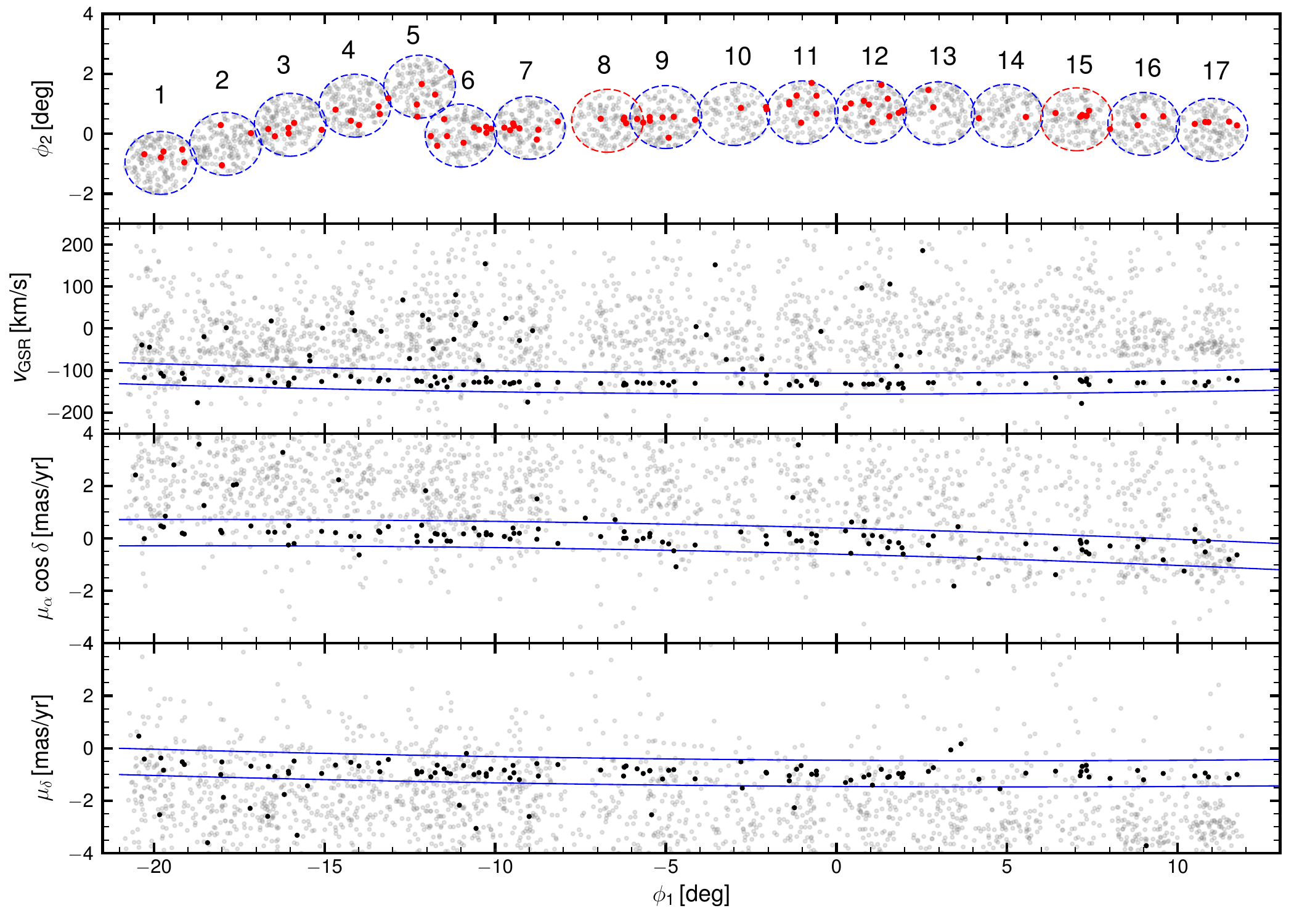}
    \caption{Selection of spectroscopic members based on radial velocity ($v_{\mathrm{GSR}}$) and proper motions ($\mu_{\alpha}\cos\delta$ and $\mu_{\delta}$) of the observed stars. In all panels, grey points show all stars observed by \SSSSS. The top panel shows the location of 17 AAT fields in stream-aligned coordinates observed in 2018; two of them (encircled in dashed red) were observed prior to \SSSSS as part of a pilot program. The red filled circles show the 96 spectroscopic members selected with the best fit track in RV and PM (blue solid lines in other three panels). The bottom three panels show the kinematic distribution of the spectroscopic sample. In all three panels, black dots show the spectroscopic sample passing the selection criteria in other two components (i.e. in between the blue lines in the other two panels). The blue lines are defined as the best fit track (see text for details) plus the width (i.e. $\pm 25$\kms in $v_{\mathrm{GSR}}$ and $\pm0.55$\masyr in both $\mu_{\alpha}\cos\delta$ and $\mu_{\delta}$). See Figure~\ref{fig:specmem} for a zoomed in version of this plot for member stars only.
    }
    \label{fig:specsel}
\end{figure*}

\begin{figure}
    \centering
    \includegraphics[width=0.47\textwidth]{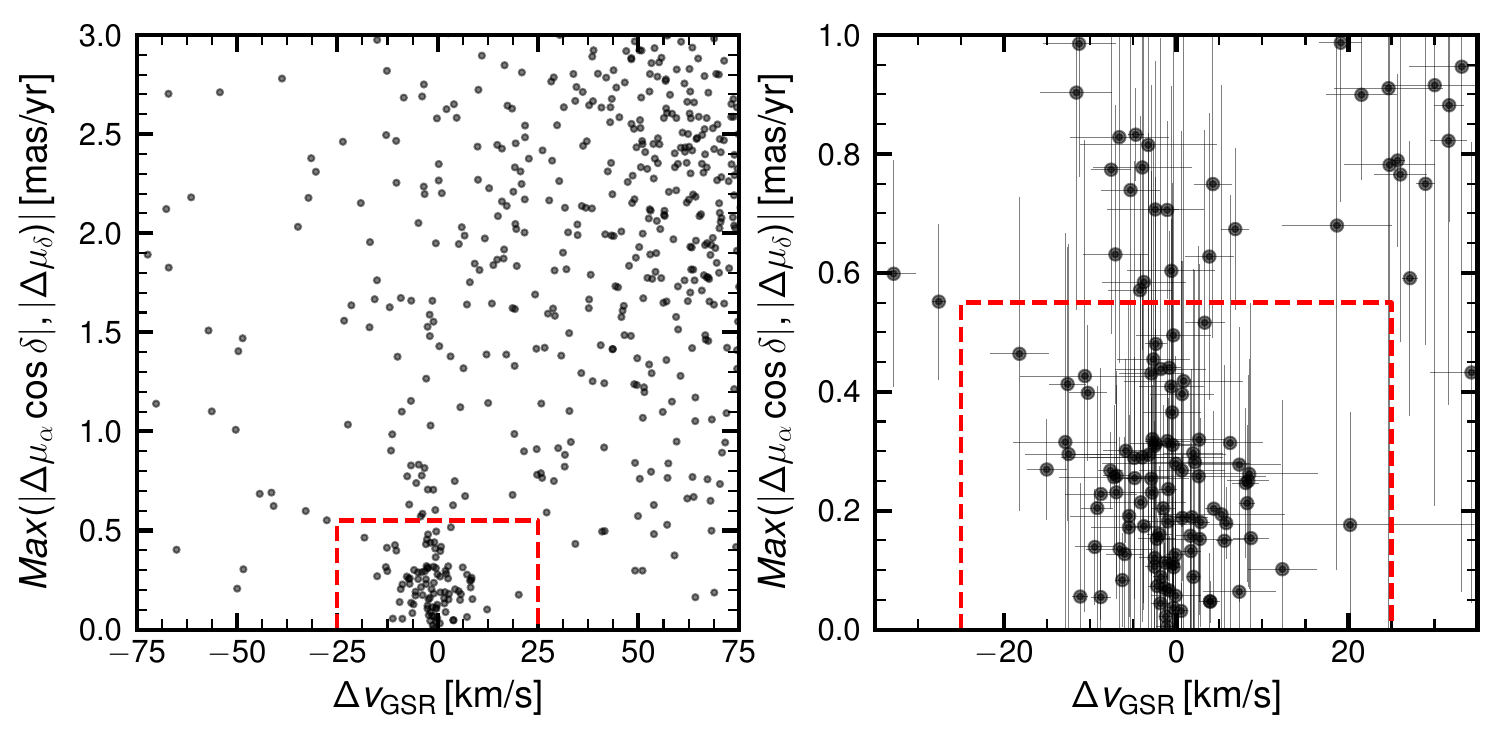}
    \caption{(left) The distance to the best fit track in RV and PM. For PM, the larger distance in either $\mu_{\alpha}\cos\delta$ or $\mu_{\delta}$ is shown. The red dashed lines show the selection width of the spectroscopic members, and stars in the red box are considered as spectroscopic members in this paper. (right) Zoomed-in version of the left panel with uncertainties shown. 
    }
    \label{fig:specsel2}
\end{figure}

\begin{figure*}
    \centering
    \includegraphics[width=0.8\textwidth]{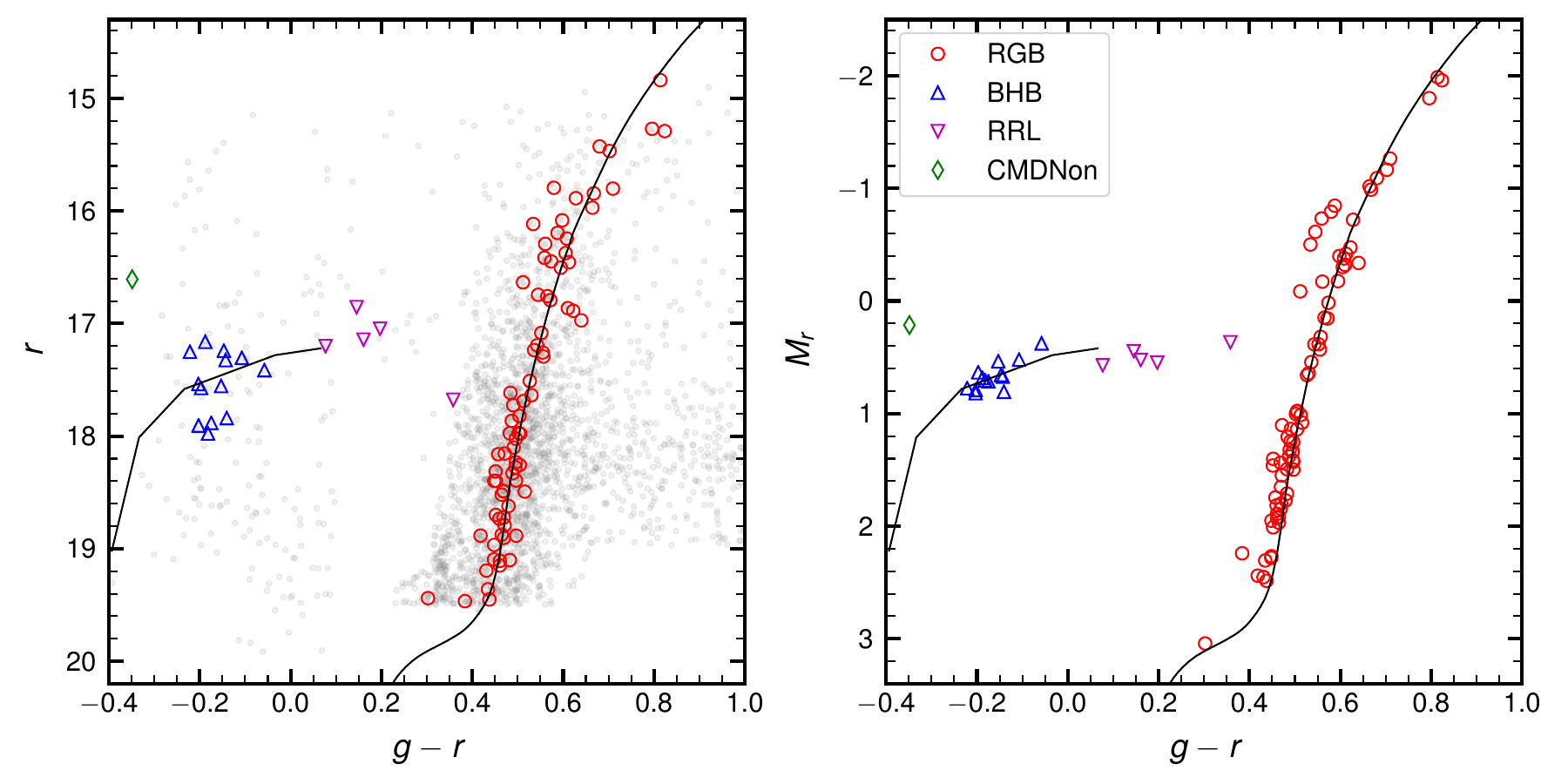}
    \caption{(left) Color-magnitude diagram of the spectroscopic stream members. While the member stars are selected kinematically (i.e. Figure~\ref{fig:specsel}), most of them can be well described by a stellar isochrone at distance modulus $m-M$ = 16.8. The BHB isochrone is taken from the globular cluster M92~\citep{Clem2006PhDT,Belokurov:2007} and the RGB isochrone is taken from the Dartmouth Stellar Evolution Database \citep{Dotter:2008} with parameters detailed in Section \ref{sec:isochrone}. Grey dots show all the stars observed in the 17 AAT fields and different symbols show member stars in different stellar populations, including red giant branch (RGB) stars, blue horizontal branch (BHB) stars, and RR Lyraes (RRL). We also note one CMD non-member (CMDNon), which has kinematic properties consistent with being a member star.
    (right) HR diagram of the same spectroscopic members corrected for the $\phi_1$ dependent distance as measured in Section~\ref{sec:distance}. 
    With distance correction, both the horizontal branch sequence and red giant branch sequence become significantly tighter. A group of Asymptotic Giant Branch (AGB) stars also become visible at $M_r\sim-0.75$. 
    }
    \label{fig:cmd}
\end{figure*}

\begin{figure*}
    \centering
    \includegraphics[width=0.89\textwidth]{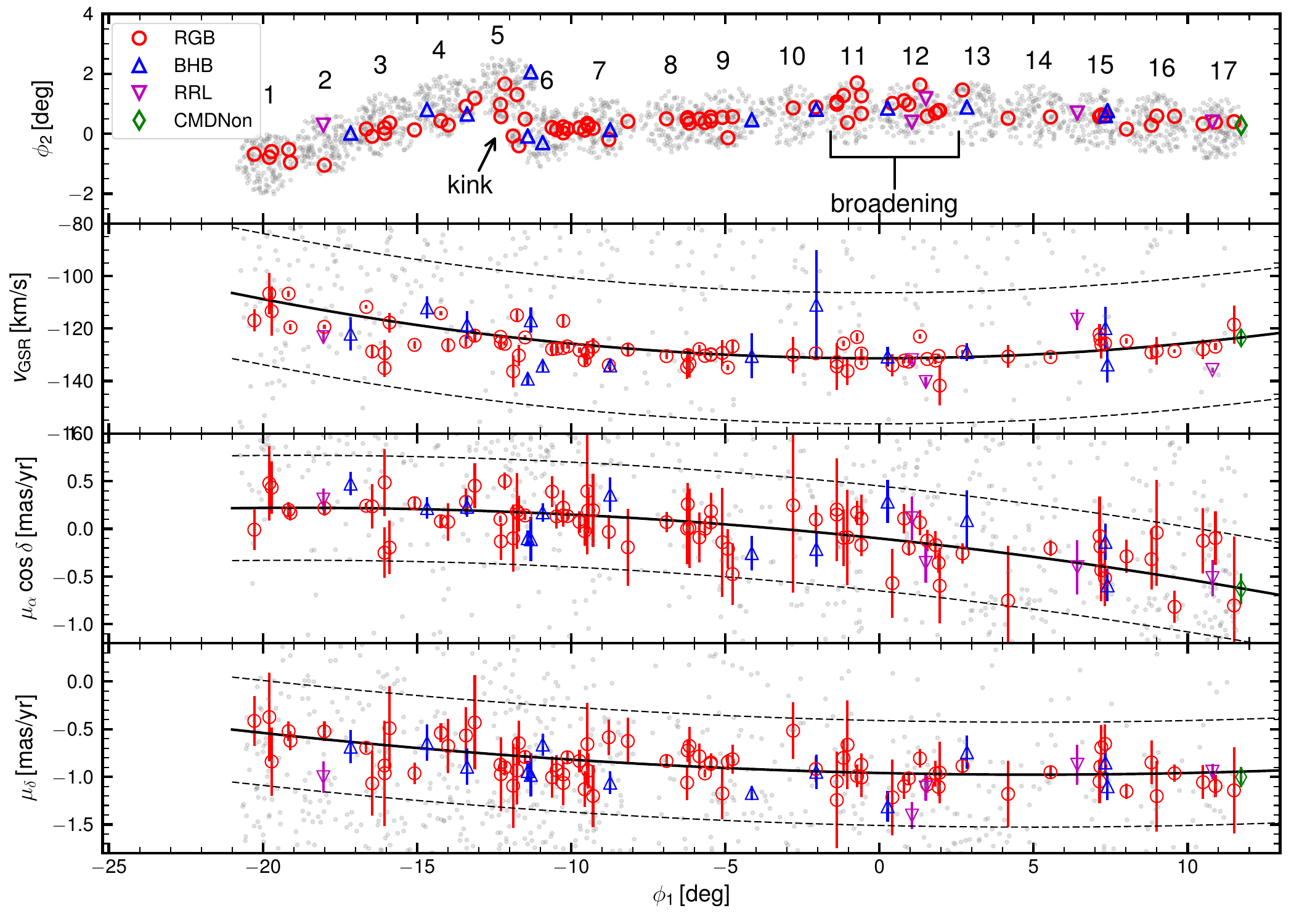}
    \caption{Zoom-in of Figure~\ref{fig:specsel} with spectroscopic members with the same symbols as in Figure~\ref{fig:cmd}. The vertical bar in each symbol shows the uncertainties of the RV or PM measurements (many stars have RV uncertainties smaller than the size of the symbol). In the bottom three panels, the solid line shows the best-fit tracks (Eqn. 1-3) and the dashed lines show the width of the spectroscopic member selection (i.e. red dashed lines in Figure~\ref{fig:specsel2}.)
    }
    \label{fig:specmem}
\end{figure*}

We use the radial velocities (RVs) from \SSSSS and proper motions (PMs) from \gaia DR2 to determine the spectroscopic members in these two streams. Meanwhile, we find the best track in RV and PM space as a function of the stream longitude, $\phi_1$, that defines the selection criteria of the spectroscopic members. 

We first select stars with \code{good\_star} = 1 (see definition in Paper I) to ensure the quality of spectral template fit and the derived radial velocities. In addition, we use parallax ($\omega$) and parallax error ($\sigma_\omega$) from \gaia DR2, and only consider stars with 
$$\omega < \mathrm{max}(3\sigma_\omega, 0.1)$$
to exclude any nearby disk stars. We then selected the spectroscopic members through an iterative process following three steps:

1. We fit a $2^\mathrm{nd}$-order polynomial function to $v_{\mathrm{GSR}}$, $\mu_{\alpha}\cos\delta$ and $\mu_{\delta}$ for the spectroscopic members to define the kinematic track of the stream. To start the first round polynomial fit, we selected an initial sample from the high priority targets with $-140<v_{\mathrm{GSR}}/\kms\ < -120$. We also ignore all the RR Lyrae member stars in fitting $v_{\mathrm{GSR}}$ as their line-of-sight velocities vary from their true systemic velocity due to pulsation.

2. We select spectroscopic members that are within $\pm0.55$\masyr in $\mu_{\alpha}\cos\delta$ and $\mu_{\delta} $ and $\pm 25$\kms in $v_{\mathrm{GSR}}$ from the best-fit track.

3. We visually inspect the spectra and the best-fit templates for these selected spectroscopic members, and we discard any members with unreliable radial velocity measurements. We note that the cut on \code{good\_star} = 1 discarded most spectra with bad template fits, and we found that $<5$\% of the selected stars did not pass our visual inspection.

We then repeat the above procedures iteratively until there are no changes in the final spectroscopic sample.  A total of 96 spectroscopic members are identified kinematically (71 in ATLAS and 25 in Aliqa Uma, presented in Table \ref{table:atlas_members}), along with the best fit track in radial velocity and in proper motion defined as:

\begin{align}\label{eq:rvtrack}
\Track_{rv} (\kms): & &v_{\mathrm{GSR}} = -131.33 + 0.07x + 5.68x^2 \nonumber \\
\Track_{\mu,\alpha} (\masyr): & & \mu_{\alpha}\cos\delta  =  -0.10-0.34x - 0.09x^2 \nonumber \\
\Track_{\mu,\delta} (\masyr): & & \mu_{\delta}  =  -0.96-0.07x + 0.07x^2
\end{align}
\noindent where $x = \phi_1/$10\degr, with $\phi_1$ measured in degrees. 

\input{table_specmem.tex}

Figure \ref{fig:specsel} shows the spectroscopic members selected with the best-fit track. In the top panel, red filled circles indicate the final spectroscopic members selected with all three components (radial velocity and two proper motions). In each of the bottom three panels, the black dots show the candidate members selected with only the other two components, i.e. black points in the fourth panel ($\mu_{\delta}$) were selected using the track in radial velocity $v_\mathrm{GSR}$ (second panel) and $\mu_\alpha\cos\delta$ (third panel). \revise{The panels clearly show that a group of likely stream members predominantly occupy the region enclosed in solid blue lines, which are the best fit tracks defined in Eqn \ref{eq:rvtrack} plus the selection width.}

In Figure \ref{fig:specsel2}, we show the distance to the best-fit track for each star in RV and PM space. 
We note that our selection window is quite narrow with respect to the uncertainties, especially in proper motion ($\pm0.55$\masyr). This is to ensure a clean sample for further investigation in the rest of this paper. Our selection will inevitably miss some members with large proper motion uncertainties at fainter magnitude. 
However, these missed member stars likely have larger measurement uncertainties, so their absence does not significantly affect the measurements of the radial velocity and proper motion tracks.

A color-magnitude diagram (CMD) of the 96 kinematically identified spectroscopic members is shown in the left panel of Figure~\ref{fig:cmd}. 
With the kinematic selection described above, we found a total of 13 blue horizontal branch (BHB) member stars at $-0.3 < (g-r)_0 < 0.0$. In addition, five members are classified as RR Lyrae stars (RRLs) in \gaia DR2.
The majority of members are red-giant branch (RGB) stars. We note that most of the kinematically selected members are well aligned with the stellar isochrone shown as the black curves. The only exception is a blue star at $(g-r)_0\sim-0.35$ and $r_0\sim16.6$. This star deviates from the other BHBs in CMD and is marked by a green diamond. It is a CMD non-member star; however, it is kinematically consistent with other stream members (see Figure~\ref{fig:specmem}; $\phi_1\sim12\degr$). 

We now have a closer look at the spectroscopic members in Figure \ref{fig:specmem}. These 96 members are coded with different symbols by their stellar populations defined in Figure \ref{fig:cmd}. We highlight that although a $\Delta v_\mathrm{GSR}$ of 25\,\kms is used for the spectroscopic member selection (dashed line in the second panel), most of the members are very close to the RV track (black line), further confirming our robust identification of the spectroscopic members.
Despite the spatial discontinuity around $\phi_1\sim-12\degr$ (Field 5 and 6), the line-of-sight velocities and proper motions of the two streams are seamlessly connected, strongly \revise{suggesting} that these two are one single stream. 
\revise{For the remainder of the paper, we will refer the two streams as the ATLAS-Aliqa Uma stream, or the AAU stream, when discussing two streams together.}
We refer to the discontinuity feature around $\phi_1\sim-12\degr$ as a ``kink" in the rest of the paper. Furthermore, when looking at top panel of Figure~\ref{fig:specmem}, we found that the stream also displays a broader width at $-2\degr<\phi_1<2\degr$ (Field 11 and 12). Such broadening in stream width might be associated with a density variation in the stream and we investigate this further via deeper photometry in Section \ref{sec:photometry}. We refer to this feature as a ``broadening" hereafter.

\subsection{Distance Gradient from BHB and RRL}\label{sec:distance}

\begin{figure}
    \centering
    \includegraphics[width=0.49\textwidth]{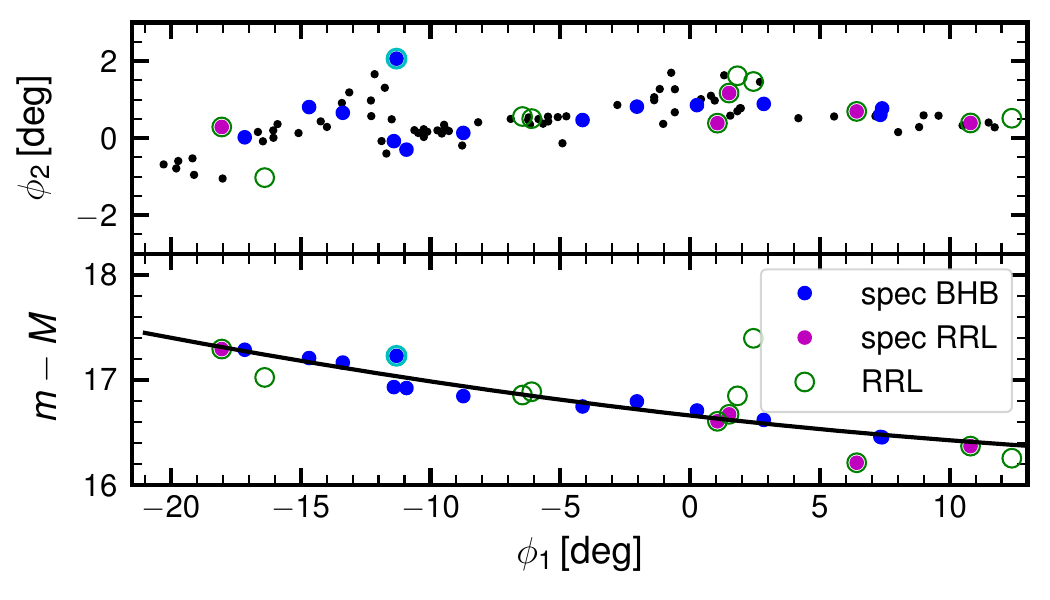}
    \caption{On sky distribution (top) and heliocentric distances (bottom) of BHBs and RRLs in the area of the streams.
    A $2^\mathrm{nd}$ order polynomial fit on distance modulus ($m-M$) of the BHBs is shown as black curve. In all panels, we also show all RRLs in the stream area with $16 < m-M < 18$ from \gaia DR2 as open circles. We note that likely not all of them are stream members. In both panels, a BHB star at $\phi_1\sim-11.5\degr$ is encircled in cyan. This BHB shows a slightly greater distance ($\Delta(m-M)\sim0.3$) than the other two BHB stars in the ATLAS stream at a similar $\phi_1$. The difference is significantly larger than the distance uncertainty of BHB stars (at 0.1--0.2 mag) and may indicate a distance spread in this area. 
    }
    \label{fig:dmgrad}
\end{figure}

In addition to the discontinuity of the two streams on the stellar density map, \citet{Shipp:2018} did not associate these two streams because their distance moduli are different by 0.5 magnitude (i.e. $m-M$ = 16.8 for ATLAS and 17.3 for Aliqa Uma). Therefore, the kinematic connection between these two streams suggests there should also be a distance gradient along these two streams. Luckily, both spectroscopically confirmed BHBs and RRLs are good distance indicators for such a study. As shown in the top panel of Figure \ref{fig:dmgrad}, BHB and RRL members are well populated along the streams. We first derive the distance modulus of each BHB member star $m-M = g - M_g$ using the $M_g$ vs $(g-r)$ relation from \citet{Belokurov2016}. Assuming the uncertainty on distance modulus for each BHB is 0.1 mag \citep{Deason:2011}, we fit the distance modulus with a second order polynomial:

\begin{equation}\label{eq:distance}
\Track_{dm}:    (m-M) = 16.66 - 0.28x + 0.045x^2
\end{equation}
\noindent where $x = \phi_1/$10\degr. We emphasize that this relation is derived using BHBs between $\phi_1\sim -17\degr$ and $\phi_1\sim7\degr$.  Extrapolation on the distance beyond these two points should be done with caution. In both panels, one BHB star around $\phi_1\sim -11.5\degr$ that is circled in cyan has a distance modulus that is 0.3 magnitudes larger than the other two BHB stars at similar $\phi_1$. This may be an indication that at the location of the ``kink" there is a distance spread, and that the Aliqa Uma stream is slightly farther than the ATLAS stream at $\phi_1\sim -11.5\degr$. This also matches with the line-of-sight velocity variation in this area as discussed later in Section \ref{sec:vdisp} and Figure \ref{fig:vdisp}.

We derived the distance using the RRL members as an independent check. To do that, we take the $M_G-\feh$ relation from \citet{Muraveva2018}, assuming a stellar metallicity of \feh = $-2.2$ (see Section \ref{sec:feh}), and $G-$band magnitude from Gaia DR2 with color-dependent extinction corrections from \citet{Babusiaux2018} and the \citet{Schlegel:1998} values of $E(B-V)$. The derived distance modulus for the confirmed RRL members are shown as magenta circles in the bottom panel of Figure~\ref{fig:dmgrad}. Four of the five spectroscopic RRL members have distances consistent with those of BHB members, and the one exception is the RRL at $\phi_1\sim 6.5\degr$. We notice that this star only has 11 transits selected for variability analysis from \gaia DR2 (\code{num\_selected\_g\_fov}=11), while the other RRL members that have over 30 transits; this might lead to an imprecise distance estimation. In addition to the spectroscopically confirmed RRL members, we checked all RRLs at $|\phi_2| < 2$ and $16 < m-M < 18$ in \gaia DR2, shown as open green circles in Figure \ref{fig:dmgrad}. While some of these RRLs are likely non-members of the streams, it is possible that two RRLs at $\phi_1\sim-6\degr$ are members of ATLAS stream, as they are at the right distance\footnote{These two RRLs are not included in the spectroscopic observations as Field 8 was observed prior to \SSSSS, so RRL candidates were not part of the target selection.}. Spectroscopic follow-up on these RRLs is necessary to confirm their membership.

\revise{Based on the distance modulus derived from BHB and RRL member stars, we confirm that not only the kinematics, but also the distance of the two streams, are seamlessly connected. Although the Aliqa Uma stream is slightly farther in distance, it is consistent with the distance gradient observed in two streams. The distances from BHB and RRL members are provided in Table \ref{table:distance}.}

\input{table_distance.tex}

\subsection{Line-of-sight Velocity Dispersion}\label{sec:vdisp}

A stream's velocity dispersion is a useful indicator of the stream's progenitor type and orbital interaction history. For example, the velocity dispersion of the Sagittarius dwarf galaxy stream is $\sim10-20$\,\kms \citep{Koposov2013,Gibbons2017} in contrast to the Palomar 5 globular cluster stream, which has a velocity dispersion of $2.1\pm0.4~\kms$~\citep{Kuzma2015MNRAS.446.3297K}. However, streams are not in dynamical equilibrium, so the dispersion cannot be directly translated to a dynamical mass for the stream progenitor.

We study the velocity dispersion along the two streams using $\Delta v_\mathrm{GSR}$, which is defined as the difference between $v_\mathrm{GSR}$ and the RV track.\footnote{RRL members are excluded in this analysis as the velocities of RRL stars varies with phase.} 
We model the $\Delta v_\mathrm{GSR}$ with a Gaussian distribution while taking into account velocity uncertainties of individual stars. The posterior on the velocity dispersion was obtained by MCMC sampling, 
similar to what has been done in kinematic studies of ultra-faint dwarf galaxies \citep[e.g.][]{Walker06,eri2}.
We use a flat prior for mean velocity and logarithmic prior (i.e. flat prior in log-space) for the velocity dispersion.  
The velocity dispersion is measured to be $4.8\pm0.4 \kms$ for the entire stream.

\begin{figure*}
    \centering
    \includegraphics[width=0.65\textwidth]{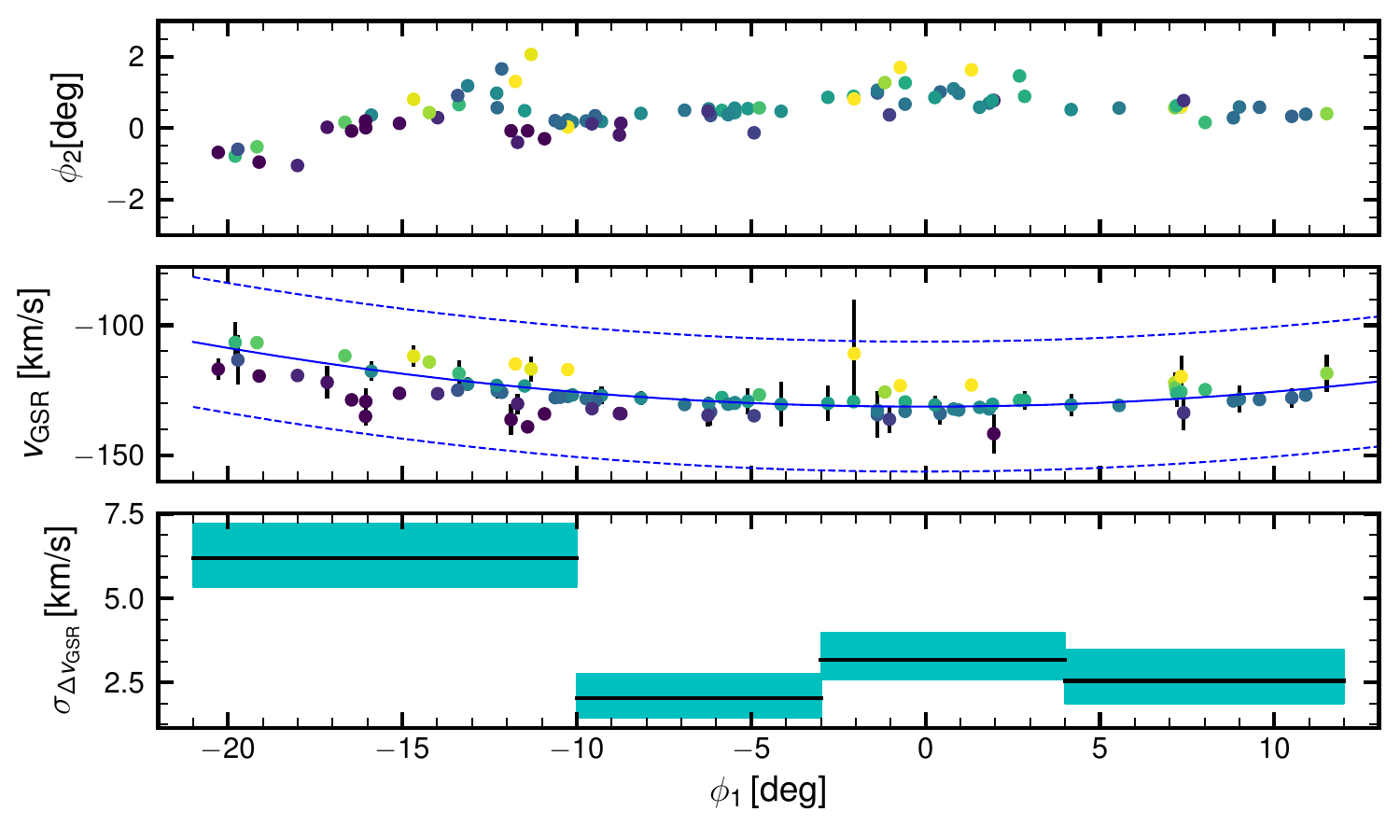}
    \includegraphics[width=0.33\textwidth]{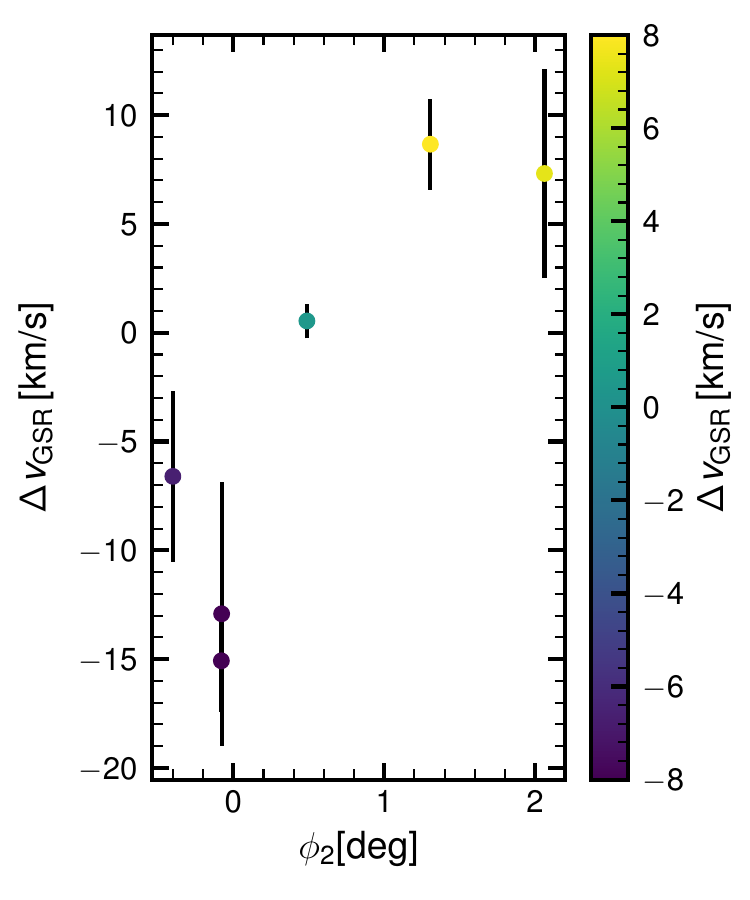}
    \caption{(Left) Velocity dispersion along the AAU stream. Top and middle panels show the spatial distribution and velocity distribution of the member stars, color-coded in $\Delta v_\mathrm{GSR}$, which is derived from the difference between $v_\mathrm{GSR}$ and the best-fit RV track (blue solid line in the middle panel). Bottom panel shows the velocity dispersion of each of four portions of the stream defined based on the stream width seen in top panel, with 1-$\sigma$ uncertainties shown as the shaded regions. (Right) $\Delta v_\mathrm{GSR}$ vs. $\phi_2$ for stars between $\phi_1 = -12\degr$ and $\phi_1 = -11\degr$. The RVs span over 20\,\kms for these six members and show a strong correlation between the position on the sky and the RVs. 
    }
    \label{fig:vdisp}
\end{figure*}

We study the variation of the velocity dispersion along the two streams in the left panel of Figure \ref{fig:vdisp}. In particular, we are interested in the velocity dispersion at the ``kink" ($\phi_1 \sim -12\degr$) and at the ``broadening"  ($\phi_1 \sim 0\degr$).
We therefore divided the streams into four portions and calculated the velocity dispersion of each portion. We found that, even with the velocity uncertainty taken into account, the dispersions around those features are indeed larger than the rest of the stream.
While the increase of the dispersion at the ``broadening" is not significant, the dispersion for Aliqa Uma is significantly larger, suggesting a severe perturbation in the past.

From the top panel of Figure \ref{fig:vdisp}, it also seems that there is a correlation between the position of the star relative the stream track on the sky and the velocity offset w.r.t. the track, i.e., $\Delta v_\mathrm{GSR}$. This is especially obvious for stars at $-12\degr \lesssim \phi_1 \lesssim -11\degr$, where the streams connect. The right panel of Figure \ref{fig:vdisp} shows a strong correlation between $\Delta v_\mathrm{GSR}$ and $\phi_2$ based on the six members in this area, which might be an imprint from an earlier perturbation. More RV measurements for stars in the connecting region between the two streams are required to understand the origin of the perturbation.

\subsection{Metallicities and Metallicity dispersion}\label{sec:feh}

As discussed in Paper I, although \code{rvspecfit} returns the stellar atmospheric parameters including metallicity of stars in \SSSSS, metallicities derived from equivalent width of Calcium triplet (CaT) lines using the relation from \citet{carrera13} show better precision when comparing with the metallicities derived from high-resolution spectroscopy, for stars with known distances such as stream members. This empirical metallicity calibration relation only applies to RGB stars with known distance, because the absolute magnitudes of the stars are required for the empirical calibration. We therefore derived the CaT metallicities for the RGB member stars using the distance relation defined in Eqn~\ref{eq:distance}.

\begin{figure*}
    \centering
    \includegraphics[width=0.46\textwidth]{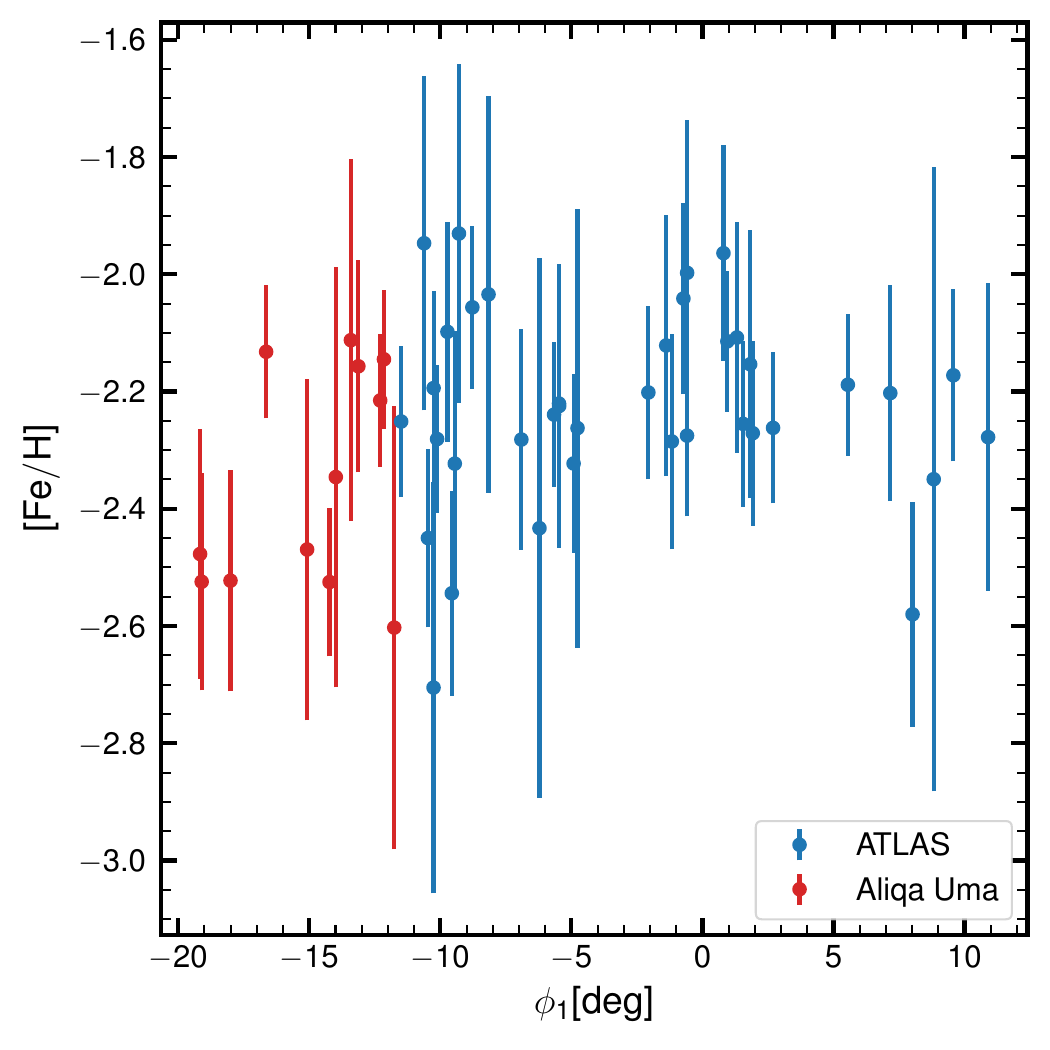}
    \includegraphics[width=0.48\textwidth]{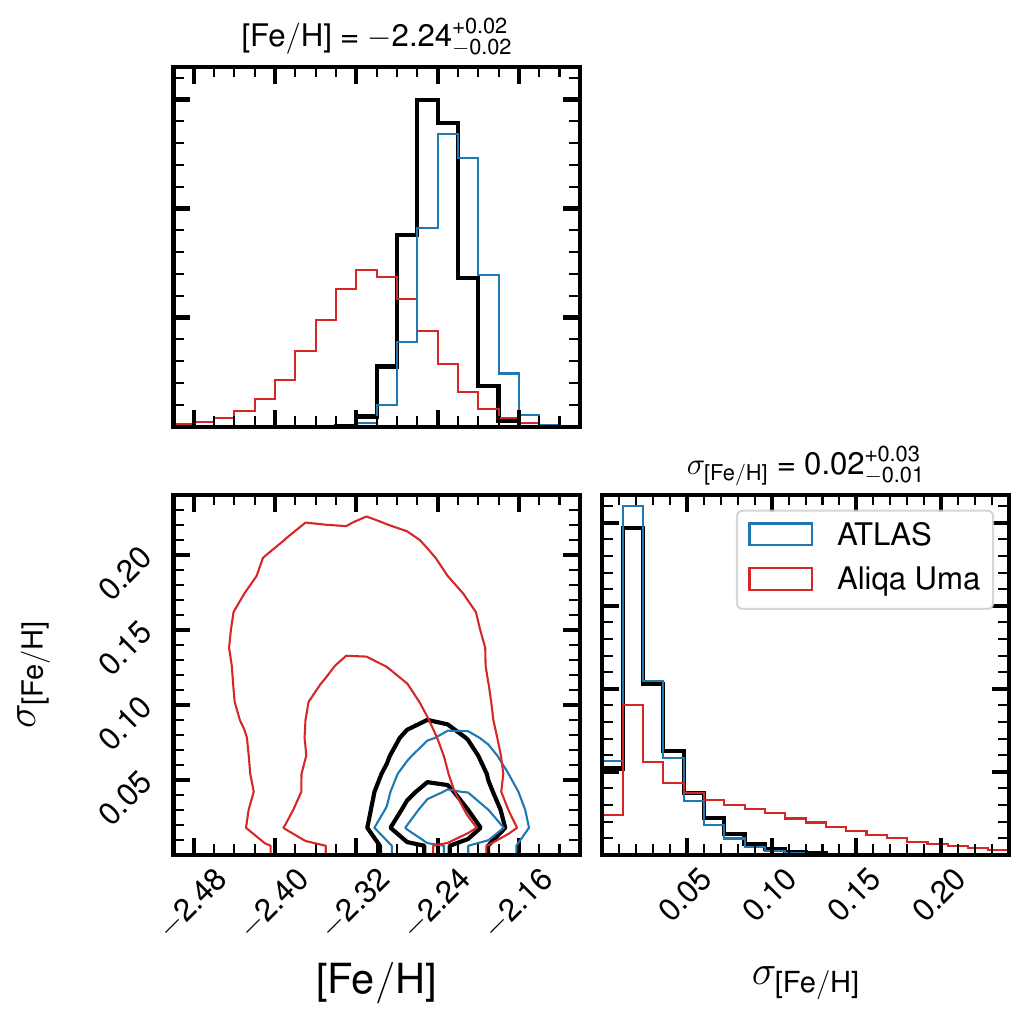}
    \caption{(left) Metallicity distribution as a function of $\phi_1$ for bright RGB member stars in ATLAS and Aliqa Uma. (right) Posterior distribution of the mean metallicity and metallicity dispersion of Aliqa Uma + ATLAS streams (black), Aliqa Uma only (red) and ATLAS only (blue). The contours correspond 68\% and 95\% confidence interval. The numbers shown in the top and right panels are the median and 1-$\sigma$ from the posterior distribution.}   
    \label{fig:feh}
\end{figure*}

The equivalent widths of the CaT lines are derived by fitting a Gaussian plus a Lorentzian function on three lines. For spectra with very low signal-to-noise ratio, the fit sometimes fails. We therefore select the RGB members with spectral S/N $>$ 8 per pixel and visually inspect the fitting quality on the equivalent widths for each individual spectrum. This results in 50 RGB members with reliable metallicity measurements which are presented in Table \ref{table:atlas_members}. The metallicities of these 50 RGBs are shown in the left panel of Figure \ref{fig:feh}. The metallicities of the two streams appear quite similar. As stars in Aliqa Uma are slightly farther away and therefore fainter, the stellar metallicities show larger scatter in smaller $\phi_1$ with larger metallicity uncertainties.

We then derive the mean metallicity and metallicity dispersion of the 
ATLAS and Aliqua Uma streams. In order to take into account the individual metallicity uncertainties in deriving the intrinsic metallicity dispersion of the system, we again applied the same method as the one used in deriving the velocity dispersion. We found a mean metallicity of $\feh = -2.24 \pm 0.02$ when combining members from both streams. The metallicity dispersion is not resolved, with an upper limit of $\sigma_{\feh} <0.07$ at 95\% confidence level. We also derive the mean metallicity and dispersion for the two streams separately (with $\feh = -2.22 \pm 0.03$ for ATLAS and $\feh = -2.30 \pm 0.06$ for Aliqa Uma).
Aliqa Uma shows a slightly lower mean metallicity but is consistent with ATLAS within 1.5-$\sigma$ uncertainty. The posterior distribution of the mean metallicity and metallicity dispersion is shown in the right panel of Figure \ref{fig:feh}.

The low metallicity dispersion suggests that the \revise{progenitor(s) of the ATLAS and Aliqa Uma streams was likely a globular cluster}. This conclusion is consistent with the thinness of the RGB of the stream members, low velocity dispersion found in the previous section, as well as the narrow width of the streams ($\lesssim 100$~pc).

\subsection{Detailed Chemical Abundances}\label{sec:abundance}

\begin{figure}
    \centering
    \includegraphics[width=\linewidth]{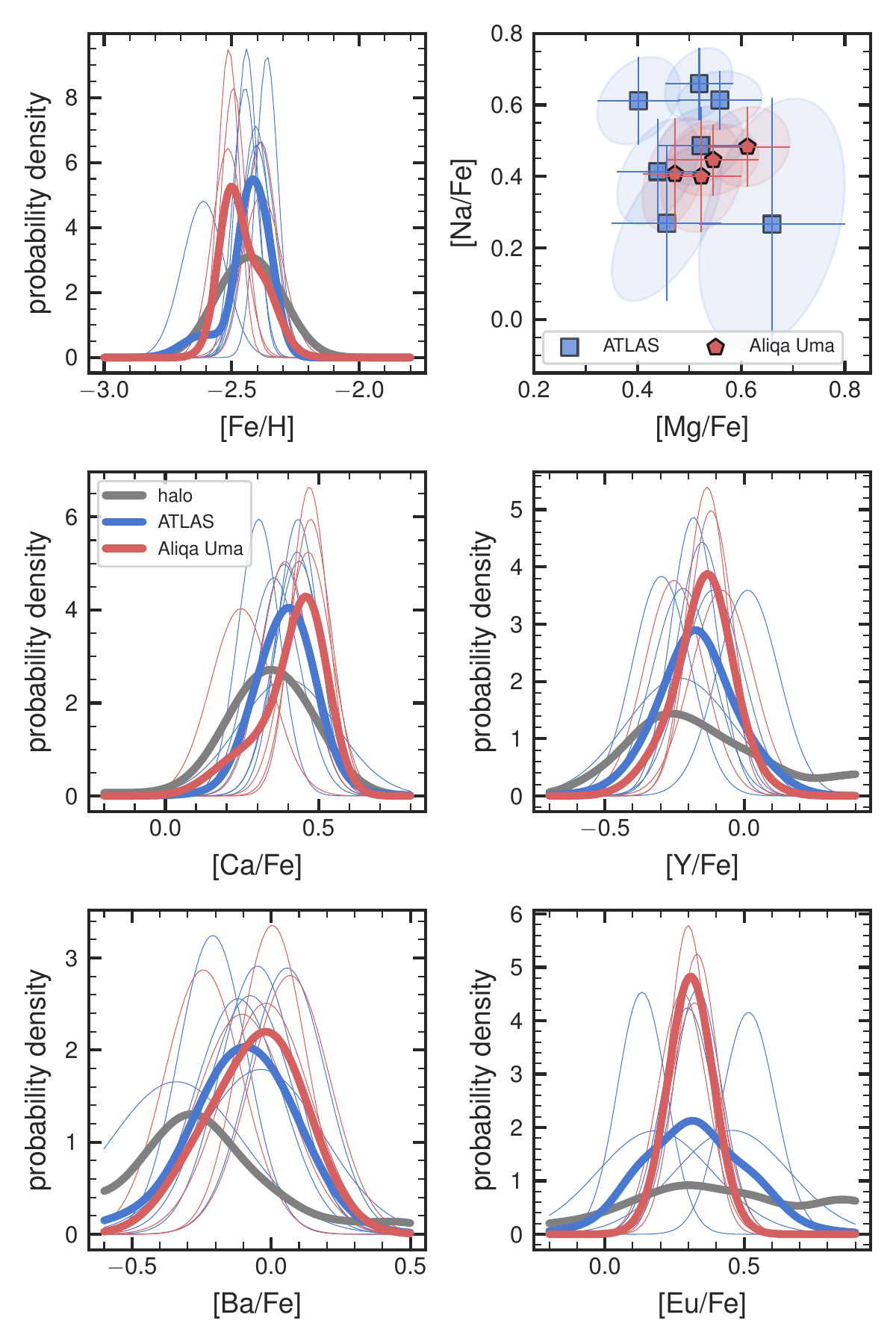}
    \caption{\textit{Top right:} Elemental abundance of ATLAS stars (blue squares) and Aliqa Uma stars (red pentagons).
    The error bars and shaded ovals indicate 1-$\sigma$ errors propagating all stellar parameter uncertainties, including correlations in [Na/Fe] and [Mg/Fe].
    \textit{Other panels:} distribution of abundance ratios in ATLAS (blue) vs Aliqa Uma (red).
    Each individual star's abundance measurement and error are treated as a Gaussian and shown as a thin colored line.
    The sum of these PDFs is indicated as a thick line.
    The thick grey line is a comparison sample of Milky Way halo stars with $-2.55 < \mbox{[Fe/H]} < -2.3$ \citep{jinabase}, a range chosen to match the MDF of ATLAS and Aliqa Uma.
    0.1 dex errors have been assumed for the halo sample.
    }
    \label{fig:hrs}
\end{figure}

In additional to the AAT observations, \SSSSS has also been collecting high-resolution $R \sim 30,000$ spectroscopy on the brightest RGB stream member stars using larger aperture telescopes. Details on the observations and abundance analysis is discussed in \citet{Ji2020b}. Here we focus on a few elements that support our claim that the ATLAS and Aliqa Uma streams share a common origin. Seven stars in ATLAS and five stars in Aliqa Uma were observed with Magellan/MIKE \citep{Bernstein03}, producing spectra with a typical S/N of 20 per pixel in the blue and 40 per pixel in the red.
In Figure~\ref{fig:hrs} we show the abundance distributions for several elements. One Aliqa Uma star has especially low S/N and is thus missing from many panels. Each individual star is plotted as a thin Gaussian with its own mean and standard deviation. ATLAS and Aliqa Uma stars are shown in blue and red, respectively. The total distribution, found by summing the individual distributions, is plotted using thick blue and red lines. It is clear that both the iron abundance, and the [X/Fe] ratios for the other elements, are essentially identical between ATLAS and Aliqa Uma. Similar convergence is seen for ${\sim}10$ additional elements not shown here \citet{Ji2020b}.
In general the abundance scatter is smaller than expected from halo stars of similar metallicity (thick grey lines), which is most clear from the neutron-capture elements (Y, Ba, Eu).

Globular clusters exhibit characteristic element anticorrelations between stars, which we do not expect to detect in the two streams given our abundance uncertainties. Given the available elements and precisions, the strongest anticorrelation we expect is between [Na/Fe] and [Mg/Fe], shown in the top-right panel of Figure~\ref{fig:hrs}. In some globular clusters, a 0.1 dex decrease in [Mg/Fe] corresponds to a 0.4 dex increase in [Na/Fe], though the extent of Mg depletion varies from cluster to cluster \citep[e.g.,][]{BastianLardo2018}. Given the uncertainties in both [Mg/Fe] and [Na/Fe], we would not expect to clearly detect this signature.

Combining the ATLAS and Aliqa Uma stars, the mean metallicity is $-2.38 \pm 0.03$ dex with 95\% confidence upper limit on the dispersion of 0.12 dex. The mean metallicity is lower than the CaT values, but within the \citet{carrera13} calibration systematic uncertainty of 0.16 dex. \revise{Individually, the ATLAS and Aliqa Uma streams have identical mean metallicities of $\mbox{[Fe/H]} = -2.36^{+0.05}_{-0.06}$ and $-2.39^{+0.06}_{-0.05}$, respectively.} 

\section{Stream properties from {\emph GAIA} DR2 and DES DR1}\label{sec:photometry}

In order to expand our study of the AAU stream beyond the spectroscopic observations, we proceed to an analysis of the photometric and astrometric only datasets from DES DR1, PS1 DR1 and \gaia DR2, which allows us to probe the stream beyond the footprint coverage and depth of \SSSSS.

\subsection{Isochrone Model}\label{sec:isochrone}

As a first step, we proceed to determining the DES color-magnitude diagram distribution of stream members. In Section~\ref{sec:member} we have shown that the spectroscopic members line up extremely well on the RGB. In order to map the stream fully we need an isochrone model that suits both the main-sequence and RGB stars in the stream. 

\begin{figure}
    \centering
    \includegraphics{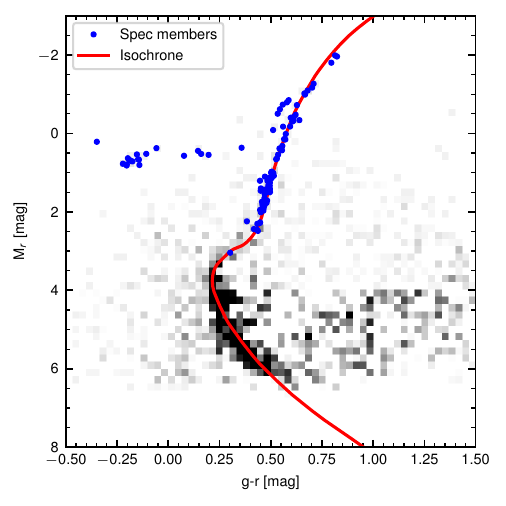}
    \caption{The color-absolute magnitude diagram of the AAU stream from the combination of photometric and spectroscopic datasets. The greyscale image shows the background subtracted Hess diagram of the ATLAS stream for the area $-11.5<\phi_1<10$ and within 0.25 deg of the approximate stream track on the sky (Eq.~\ref{eq:track_sky}). The photometric only Hess diagram is complemented by the spectroscopic members of the AAU shown in blue circles. The red curve is the best isochrone that matches both the main sequence stars from the deep photometric data as well as the giants from the spectroscopic sample (see main text for details). The magnitudes in this plot have been corrected by the distance modulus as a function of $\phi_1$ determined in Section~\ref{sec:distance}.}
    \label{fig:combcmd}
\end{figure}

To find that model, we take an approximate stream track from the spectroscopic stream members 
\begin{equation}
\Track_{\phi_2}(\phi_1) = \Delta -0.5((\phi_1-3)/10)^2 \ \ \mathrm{degrees}
\label{eq:track_sky}
\end{equation}
where $\phi_1$ is measured in degrees and
where $\Delta=0.75$ for $\phi_1>-11.5$\degr and $\Delta=1.5$ otherwise.
Then we construct the background subtracted Hess diagram for the region $|\phi_2-\Track_{\phi_2}(\phi_1)|<0.25\degr$ around the track, using 
the two bands outside the stream region $1\degr<|\phi_2-\Track_{\phi_2}(\phi_1)|<2\degr$ as a background. We also correct the magnitudes for the distance modulus changes along the stream as measured in Section~\ref{sec:distance}. The resulting Hess diagram is shown in Figure~\ref{fig:combcmd}, with the absolute $r$ magnitude and $(g-r)$ color for spectroscopic members overplotted. The figure clearly shows a main sequence turn-off (MSTO) that smoothly transitions into the red giant branch that is well traced by the spectroscopic members. 

We attempted to identify the best isochrone describing the stellar population of the stream using various isochrone sets, such as PARSEC \citep{Bressan:2012}, Dartmouth \citep{Dotter:2008} and MIST \citep{Dotter2016,Choi2016}. However, we were not able to find one that could well fit the extremely precise measurement shown on Figure~\ref{fig:combcmd}. We therefore systematically searched for an isochrone that could match the data with the help of shifts in color and magnitude ($g-r$, $r$). The best match was found to be a Dartmouth isochrone with parameters [Fe/H]=$-$1.99 , [$\alpha$/Fe]=0.4, $Y= 0.4$, Age=$11.5$\,Gyr\footnote{Filename for the best match is \code{DECam/fehm20afep4y40} }
that needed to be shifted by $0.143,0.188$ in $g,r$, respectively. We remark that this shift is mostly in absolute magnitude, as the color shift is only $\sim$0.04. This implies a possible mismatch in the BHB distance and MSTO distance at 0.1 mag level.
This isochrone is shown by a red curve on the Figure. We note that the isochrone match is performed to get an isochrone track for the density map construction in next Section; the isochrone parameters such as metallicity, $\alpha$-abundance and age may not be best estimates of the properties the AAU stream, since shifts in magnitude and color are applied to get the best matching isochrone.

\subsection{Probable stream members with Gaia}
\label{sec:gaia_selection}

We start by constructing a map of the stellar streams using the \gaia astrometric data combined with accurate ground-based photometry.  For this we will rely on the results from Section~\ref{sec:spec}, where we determined the track of the stream in proper motion and distance space, as well as on the stream isochrone, established in the previous section. 
\begin{figure*}
    \centering
    \includegraphics[width=0.99\textwidth]{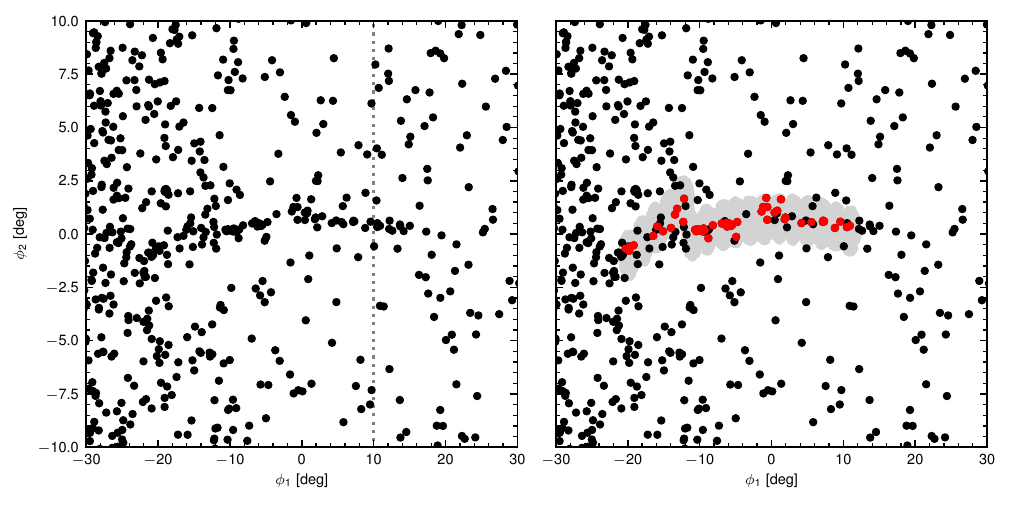}
    \caption{The distribution of stars on the sky in the region near the AAU stream, selected using astrometry from \gaia DR2 and photometry from PS1 and DES DR1 (identical for both left and right panels). As comparison, on the right panel we also show in grey the location of the \SSSSS fields and mark in red the stars among the selected ones that are spectroscopic members identified in Section~\ref{sec:member}. The dashed line at $\phi_1=10$ shows the boundary of the DES footprint. To the left of the line we use the photometry from DES, and to the right of the region we use the PS1 photometry. We remark that the stream is clearly extending well beyond our spectroscopic coverage to $\phi_1\sim20\degr$, and potentially to $\phi_1<-20\degr$. 
    }
    \label{fig:gaia_map}
\end{figure*}

Our primary astrometric selection based on proper motions and parallax is: 
$$|\mu_\alpha-\Track_{\mu,\alpha}(\phi_1)| < 0.2 + 2 \sigma_{\mu,\alpha}$$
$$|\mu_\delta-\Track_{\mu,\delta}(\phi_1)| < 0.2 + 2 \sigma_{\mu,\delta}$$
$$\omega < 0.05 + 2.5 \sigma_\omega$$
where the proper motion is in $\masyr$.
We then combine it with the color-magnitude selection based on photometric data from different ground-based imaging surveys. 
As the DES DR1 data is only available for the region of the stream with $\phi_1<10\degr$, we were required to use photometric measurements from other surveys in the region $\phi_1>10$. We decided to rely on the PS1 photometry provided in the \code{MeanObject} table. The DES and PS1 color-magnitude selection is $M_r(\phi_1)<2$ and $|g-r-I_{g-r}(r-\Track_{dm}(\phi_1))|<0.02$ where 
$I_{g-r}(M_r)$ is the best isochrone predicted color for a given $M_r$ as described in the previous section. Furthermore, we used simple linear corrections determined from a DES/PS1 overlap to convert the DES isochrone into the PS1 photometric system ($ g_\mathrm{PS1} =  g_\mathrm{DES} - 0.05 (g_\mathrm{DES}-r_\mathrm{DES})$, $r_\mathrm{PS1} = r_\mathrm{DES} + 0.08 (g_\mathrm{DES} - r_\mathrm{DES})$).
 
Figure~\ref{fig:gaia_map} shows the density of likely stream members according to the combined astrometric and color-magnitude selection. We also mark the stars that are identified as spectroscopic members in red (right panel). We see that the \gaia selected stars clearly show both the ATLAS and Aliqa Uma streams. We also see that the spectroscopic members trace the streams well, without missing significant parts. However, there is a somewhat overdense area at $-25\degr<\phi_1<-10\degr$ below the Aliqa Uma stream, where there could be more unidentified members. Also the \gaia selected stars seem to show a possible ``spur" --- stars offset from the main stream track --- at $\phi_1=-10\degr$, $\phi_2\sim 2\degr$ coming out of the continuation of the Aliqa Uma stream, and for which we could be possibly missing some members. 
Furthermore, the data suggests that the stream extends significantly further than indicated by the DES data, by  $>10$ degrees up to $\phi_1\sim 20\degr$, supporting what was seen in PS1 data by \citet{Bernard:2016}.

\subsection{Spatial density map with DES}
\label{sec:DES_phot}

Having  used  the \gaia data to map the brightest members in the AAU stream, we now proceed to use the deep DES data alone (which extends below the MSTO of the stream) to extract the stream track and density variations.
To select only point sources from DES we apply the following two selections.

\begin{align}
\left\lvert 
{\frac{S_G}{SE_G^2} +
\frac{S_R}{SE_R^2}}\right\rvert\times \left({\frac{1}{SE_G^2} +
\frac{1}{SE_R^2} }\right)^{-1}
< 0.003
\end{align} 
\begin{align}
 | r - i - 0.04 - 0.4 \,(g - r  - 0.25)|<0.1
  \end{align}
where $S_G,S_R, SE_G,SE_R$ are the 
{\tt SPREAD\_MODEL} quantities in $g$ and $r$ filters and their uncertainties respectively. 
The first selection is a morphological selection \citep{Desai2012,Koposov2015}, while the latter is a stellar locus selection.

To proceed with the mapping we use several ingredients that we have determined in previous sections, such as the isochrone model of the stream determined in Section~\ref{sec:isochrone} and the distance track determined in Section~\ref{sec:distance}. 
With this we can construct the probability  distribution of stream members in CMD space as a function of $\phi_1$, ${\mathcal P}(g-r,r |\phi_1, {\rm stream})$. We can also construct the color-magnitude distribution model of the background ${\mathcal P}(g-r,r|{\rm background})$ (we assume that the background color-magnitude distribution does not depend on $\phi_1$). With these two probability distributions we can use the matched filter approach from~\citet{Rockosi2002} where we weight each star by the ratio of ${\mathcal P}(g-r,r |\phi_1, {\rm stream})$ and
${\mathcal P}(g-r,r |\phi_1, {\rm background})$. We however adopted instead the binary matched filter method from \citet{Erkal:2017}, in which a weight of one is assigned to stars with ${\mathcal P}(g-r,r |\phi_1, {\rm stream})/{\mathcal P}(g-r,r |{\rm background})> T$ where $T$ is the threshold chosen to maximize the signal to noise of the map, and zero otherwise. The advantage of the latter approach is that it produces a map with Poisson distributed values. 

When applying the matched filter to the data we split the considered $\phi_1$ range into 100 intervals, and for each interval of $\phi_1$ we compute an optimal matched filter mask. The $\phi_1$ range needs to be split because the best CMD mask will change as the stream distance changes. This should produce the optimal map of the stream, with the only caveat being that any large scale density variations along $\phi_1$ will be somewhat modulated by the changing color-magnitude filter along $\phi_1$. 

\begin{figure*}
    \centering
    \includegraphics[width=0.99\textwidth]{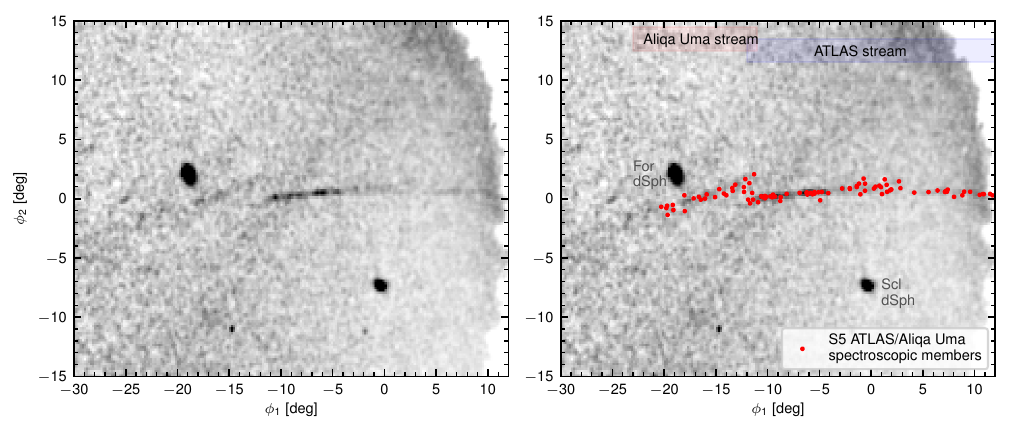}
    \caption{The stellar density of stars in DES DR1, selected using a $\phi_1$-dependent matched filter, that relies on the distance track as determined in Section \ref{sec:distance} (left). The density has been computed in square bins of $0.2\degr\times0.2\degr$ and convolved with Epanechnikov kernel with the width of 3 pixels. Each column of the image was normalized by the mean background value at a given $\phi_1$ to correct for the background density variation along $\phi_1$. The greyscale is linear with black corresponding to value of 4 and white to value of 0.2. The right panel shows the same stellar density map with the spectroscopic members overplotted on the stellar stream. The prominent overdensities visible on the map at $\phi_1,\phi_2=(0\degr,-7\degr)$ and $(-19\degr,2\degr)$ are Sculptor and Fornax dwarf spheroidals respectively, that are located at distances significantly farther than the streams.
    }
    \label{fig:des_map}
\end{figure*}

Figure~\ref{fig:des_map} shows the matched filter map of the streams. The image has also been smoothed with a rectangular Epanechnikov kernel with a width of 3 pixels and normalized along columns to have the same mean to correct for variable stellar density along the field. Both panels show the same data, but on the right panel we also overplot the location of spectroscopic members, identified in Section~\ref{sec:member}. The left panel clearly shows two streams that look unconnected. However, we see that the spectroscopic members show a bridge connecting the streams. This suggests that in fact the area near $\phi_1\sim -12\degr $ between two streams likely has some low-surface brightness stellar spray that is only detectable with spectroscopy. 
Another major feature visible on the map is density variations. We notice multiple such features. The bright part of the ATLAS stream in the range $-12\degr<\phi_1<-5\degr$ shows small-scale ($\sim 1\degr$) density oscillations, and there is an extreme density drop near $\phi_1\sim 3\degr$. We will discuss this feature later, but we remark that this density drop is accompanied by the significant broadening of spectroscopic members in $\phi_2$. It may also be noticed that the stream to the right of the gap at $\phi_1\sim 3\degr$ is shifted down in $\phi_2$ with respect to the stream on the left (we confirm this shift with stream track measurements at the end of this Section).

To fully characterize multiple observed features in the stream we need to construct a stream model.
We follow the generative stream model approach presented in~\citet{Erkal:2017} and \citet{Koposov2019} based on using natural cubic splines with different numbers of knots to describe various stream properties, such as stream density, width, track and background. 
Specifically, we use a model implemented in the \code{STAN} programming language \citep{Carpenter2017} that is almost identical to the one published in \citet{Koposov2019}. 
This implementation allows us to perform the sampling of the posterior using a technique that is highly efficient in high-dimensional spaces, Hamiltonian Monte Carlo \citep{Neal2012,Betancourt2017}, and specifically its adaptive version called No-U-Turn Sampler \citep{Hoffman2011}. 

Our model fits for the ${\mathcal B}(\phi_1)$, ${\mathcal B_1}(\phi_1)$ $ {\mathcal
  B_2} (\phi_1)$, ${\mathcal I}(\phi_1)$, ${\mathcal S}(\phi_1)$, ${\Phi}_2(\phi_1)$ which are
the splines for the logarithm of the background density, the slope of log-background
across the stream, the quadratic term for the log-background, the logarithm of stream's
central stellar density, the logarithm of the stream width, and stream track on the sky, respectively.
The parameters of the model are the values of the spline
at the spline nodes/knots. 
The profile of the stream is assumed to be Gaussian along $\phi_2$. 
More details of the implementations are described in \citet{Koposov2019}. The data that we model is the binned stellar density maps of matched filter selected stars (as described above). The bin-size is $0.2$\degr\ in $\phi_1$ and $0.05$\degr\ in the $\phi_2$ direction. We assume that the number counts in each pixel is a Poisson variate with the rate parameter determined by our density model. We decided to model the ATLAS and Aliqa Uma streams separately by focusing on the range of {$-21\degr<\phi_1<-10\degr$} for Aliqa Uma and {$-13\degr<\phi_1<10\degr$} range for the ATLAS stream. As opposed to \citet{Erkal:2017}, but similarly to \citet{Koposov2019} we use equidistant spline knots. We determine the best number of knots $k_{\Phi,2},k_{\mathcal I},k_ {\mathcal B},k_{{\mathcal B,1}},k_{{\mathcal B},2}$ for each spline by running Bayesian optimization \citep{Gonzalez2016,Gpyopt2016} of the cross-validated (K=3) log-likelihood function with respect to the vector of number of knots. The cross-validation was performed by randomly assigning pixels on the sky to one of the 3 groups. We only manually fix the number of knots for the stream width spline to 3 for Aliqa Uma and 15 for ATLAS. The optimization leads to $k_{\Phi,2},k_{\mathcal I},k_ {\mathcal B},k_{{\mathcal B,1}},k_{{\mathcal B},2}=(10,17,28,11,3)$
nodes for the stream track, stream surface brightness, log-background, background slope and background quadratic slope for the ATLAS stream and $(5,5,3,6,3)$ for the Aliqa Uma stream respectively. The spline models are then fitted to the data, with posterior samples computed using 12 independent chains running for 2000 iterations with the first half discarded. All the chains that we use show the satisfactory value of the Gelman-Rubin convergence diagnostic \citep{Gelman1992,Gelman2013} of $\hat{R}<1.1$. 

\begin{figure}
    \centering
    \includegraphics{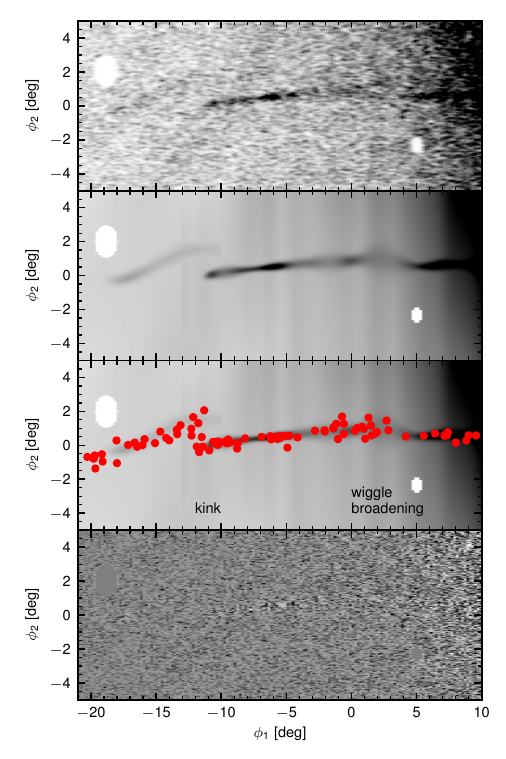}
    \caption{The results from modeling the density of the streams with DES DR1 photometry data. {\it Top panel:} The density of stream stars selected using the matched-filter mask. The panel relies on the same data as used in Figure~\ref{fig:des_map}, but shows only the modeled region with the same binning as used for the model fitting. {\it Second panel:} The maximum-a-posteriori (MAP) model of the data shown in the top panel. The model is a combination of two separate models, one for Aliqa Uma and another for the ATLAS stream.
    {\it Third panel:} The model with the spectroscopic members overplotted. {\it Bottom panel:} The residual density map showing the observed density minus the MAP model of the density.
    Two circle-shaped gaps seen in the data and models in all panels at $(\phi_1,\phi_2)=(-19\degr,2\degr)$ and $(\phi_1,\phi_2)=(5\degr,-2.5\degr)$ show the masked regions around Fornax dwarf spheroidal and NGC 288 globular cluster, respectively.
    }
    \label{fig:modelcomparison}
\end{figure}

\begin{figure}
    \centering
    \includegraphics{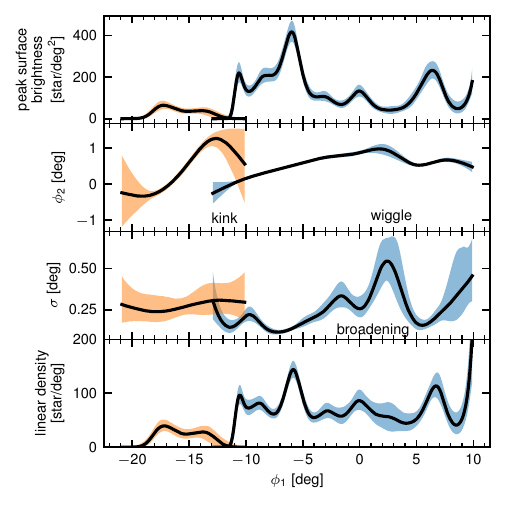}
    \caption{Measurement of stellar stream parameters as function of the position along the stream from modeling the density maps based on DES DR1 photometry. From top to bottom are stream surface brightness, stream track, stream width, and linear density, respectively. The shaded area shows the 1-sigma uncertainty from the posterior distribution. We remark that the stream densities shown here are for the optimal matched filter selection from DES data that is smoothly changing as a function of $\phi_1$, since the stream distance changes. Because of that, large scale density trends (tens of degrees) are not representative of the overall stream surface brightness changes, while small scales robustly show over- and under-densities.}
    \label{fig:posterior}
\end{figure}

The results of the model are shown in Figures~\ref{fig:modelcomparison} and \ref{fig:posterior}. Figure~\ref{fig:modelcomparison} shows the best-fit model (second panel from the top) as it compares to the data (top panel) and the spectroscopic member distribution (third panel from the top). We also show that the model residuals are negligible (bottom panel).  The key feature that we want to highlight is that at $\phi_1\sim 3\degr$, and possibly $\phi_1\sim -2\degr$, the model noticeably broadens, and simultaneously the spectroscopic members also show significantly broader distribution. We emphasize that the spectroscopic members are sampling much shallower data than what was used in the modeling, and therefore provide an independent assessment on these features.  We also notice that our model does not detect an apparent connection between two streams, but the presence of spectroscopic members in between the two streams at $\phi_1\sim -12\degr$ suggests that there is a low surface brightness spray of stars between the streams. 

To better assess the behavior of the streams captured by our model it is also informative to look at the extracted stream parameters shown in Figure~\ref{fig:posterior}. Here we show the stream surface brightness, on-sky track, stream width and linear density for both streams. This plot confirms several features that we have remarked on previously.  The first one is we see the strong stream surface brightness variations in ATLAS. The surface brightness changes by a factor of almost 10 from one position within the main part of the stream to another. Unsurprisingly, as clearly seen in Figure~\ref{fig:des_map}, the surface brightness of the Aliqa Uma stream is also significantly lower than that of ATLAS. 
The tracks of two streams show that the Aliqa Uma stream is offset and somewhat tilted with respect to the ATLAS stream. The extracted tracks also confirm that the ATLAS stream shows a clear shift in the track at $\phi_1\sim 3\degr$ of $\sim 0.2$\degr, which we refer to as a ``wiggle". 
This shift also coincides with the observed stream broadening which is clearly visible in the stream width track in the third panel of Figure~\ref{fig:posterior} as well as in Figure~\ref{fig:modelcomparison}. We also notice that there is possibly another broadening at $\phi_1 \sim -2\degr$, followed by narrowing near $\phi_1\sim 0\degr$. The distribution of spectroscopic members seems to support this picture, but deeper data are needed to confirm the observed behavior. There is also a well defined overdensity in the stream at $\phi_1\sim 6\degr-7\degr$. This overdensity is apparent in both surface brightness and linear density and also seems to correspond to a very compact group of spectroscopic members seen in Figure~\ref{fig:specsel} at $(\phi_1,\phi_2)=(7,0)\degr$. Another feature seen in Figure~\ref{fig:posterior} is that the stream seems to narrow to $\sim 0.1\degr$ at its narrowest point at $\phi_1\sim -6\degr$. At this location the stream has the highest surface brightness and linear density. The Aliqa Uma stream seems to be significantly broader than the ATLAS stream. Finally, we also comment on the linear density profile. We notice that the linear density in ATLAS seems less variable than the surface brightness, suggesting that the main type of stream perturbation is stream broadening that does not affect the linear density significantly.

\section{Dynamical Modeling}\label{sec:modeling}

Equipped with measurements of the radial velocity, proper motions, distance modulus, and stream track we now fit a dynamical model to the data. In this analysis, we choose to only fit the ATLAS stream and ignore data from Aliqa Uma. This is because stream models in a smooth, time-independent Milky Way potential are not capable of reproducing the observed kink between ATLAS and Aliqa Uma. We do not attempt to separately fit Aliqa Uma since we consider this to be a perturbed part of the AAU stream. In Section \ref{sec:perturb} below we consider perturbations to our stream model from the Milky Way bar and giant molecular clouds which are known to perturb streams in the inner Galaxy \citep[e.g.][]{Amorisco2016,Erkal:2017,Pearson2017}. 

For this fit, we use the modified Lagrange Cloud stripping code \citep[mLCs][]{Gibbons:2014} which has been adapted to include the effect of the Large Magellanic Cloud (LMC) \citep{Erkal2019}. We take the parameters for the Milky Way model from \cite{McMillan:2017}; specifically, instead of using the best-fit parameters from that work, we perform our fits on 10 posterior samples of the Milky Way potential from \cite{McMillan:2017}. 
\texttt{galpot} \citep{Dehnen:1998} is used to evaluate the force from this potential but we perform the stream disruption and orbit integration using the mLCs code. We model the progenitor of ATLAS as a $2\times10^{4} \mathrm{M}_\odot$ Plummer sphere \citep{1911MNRAS..71..460P} with a scale radius of 10 pc, and this produces a stream with a similar width to ATLAS. For the LMC, motivated by the LMC mass measured in \cite{Erkal2019}, we use a Hernquist profile \citep{Hernquist:1990} with a mass of $1.5\times10^{11}\mathrm{M}_\odot$ and a scale radius of 17.13 kpc. This LMC model matches the observed rotation curve of the LMC at 8.7 kpc \citep{vanderMarel:2014}. We compute the present-day position and velocity of the LMC using its radial velocity \citep{vanderMarel:2002}, proper motions \citep{kallivayalil13}, and distance \citep{Pietrzyski:2013}.

For the data, we use the radial velocity and proper motion of the spectroscopically confirmed members from Section \ref{sec:member}. For the on-sky position, we use the stream track measured in Section \ref{sec:DES_phot}, which is more precise than using the location of the spectroscopically confirmed members. For the distance we use the polynomial for the distance measurement of Eqn. \ref{eq:distance} with its associated covariance matrix for polynomial coefficients.

We compute the likelihood of each model stream by making mock observations and comparing this with the data. The log likelihood for each data point is 

\begin{eqnarray} \label{eq:logL}
\log \mathcal{L}_i &=& -\frac{1}{2}\log\Big( 2\pi(\sigma_{\rm i,\,obs}^2+\sigma_{\rm i,\,sim}^2)\Big)\nonumber \\ &&-\frac{1}{2} \frac{(m_{\rm i,\,obs}-m_{\rm i,\,sim})^2}{\sigma_{\rm i,\,obs}^2+\sigma_{\rm i,\,sim}^2}, 
\end{eqnarray}
where $m_{\rm i,\,obs}$ is the observed value (e.g. the radial velocity of a star), $\sigma_{\rm i,\,obs}$ is the uncertainty on the observed value, $m_{\rm i,\,sim}$ is the value of the mock observation in the simulation, and $\sigma_{i,\,obs}$ is the uncertainty on the mock observation.

For the track on the sky, the data we use is the spline fit to the stream track from Section \ref{sec:DES_phot}. We fit a line using least squares to the simulated stream particles within 1.28\degr\ in $\phi_1$ of each node of the stream track to determine the sky position of the simulated stream and its associated uncertainty on the mean. The observed value at the node and its uncertainty are then compared with simulated value and its uncertainty using Eqn. \ref{eq:logL}.

For the proper motions and radial velocities, we use the measurements for each star. We fit a line to mock observations of the simulated stream within 1.26\degr\ of each star. This linear fit gives the mean and standard deviation of the mock observable at the location of the star. To compute the likelihood we then compare the observed radial velocity (proper motion) and its associated uncertainty with the velocity (proper motion) of the simulated stream at that location. We use the width of the mock observable as $\sigma_{\rm i,\,sim}$. Finally, for the distance modulus we make a mock observation of the distance and fit a quadratic over the same $\phi_1$ range as the BHBs and RRLs in ATLAS (see Figure \ref{fig:dmgrad}). We then compare this with the observed fit, accounting for the covariance in both the model and the data. 

We explore the likelihood space using the MCMC code \code{emcee} \citep{Foreman_Mackey:2013}. We stress that for each MCMC we performed, we used a fixed Milky Way potential so we are not fitting the potential but instead finding the best stream in that potential. We choose to place the progenitor of the stream at $\phi_{1,prog} = 0\degr$ and thus our free parameters are the progenitor's other coordinate on the sky ($\phi_{2,prog}$), radial velocity ($v_{r,prog}$), proper motions ($\mu_{\alpha,prog}^*,\mu_{\delta,prog}$), and distance ($d_{prog}$). We take a normally distributed prior on the distance of ($22.9\pm1$ kpc) from the measurement in \cite{Shipp:2018}. For the proper motions and radial velocity we use uniform priors which are broad, $|\mu_\alpha^*|<10$\masyr,$|\mu_\delta|<10$\masyr, and $|v_r| < 500$\kms. We give a uniform prior on $\phi_{2,prog}$ with $-2\degr < \phi_{2,prog} < 2\degr$. We use 100 walkers for 2000 steps with a burn-in of 1000 steps. We note that for all of the subsequent analysis in this work, we only use the Milky Way realization from \cite{McMillan:2017} which gave the best-fit to ATLAS stream.

\begin{figure}
    \centering
    \includegraphics[width=0.49\textwidth]{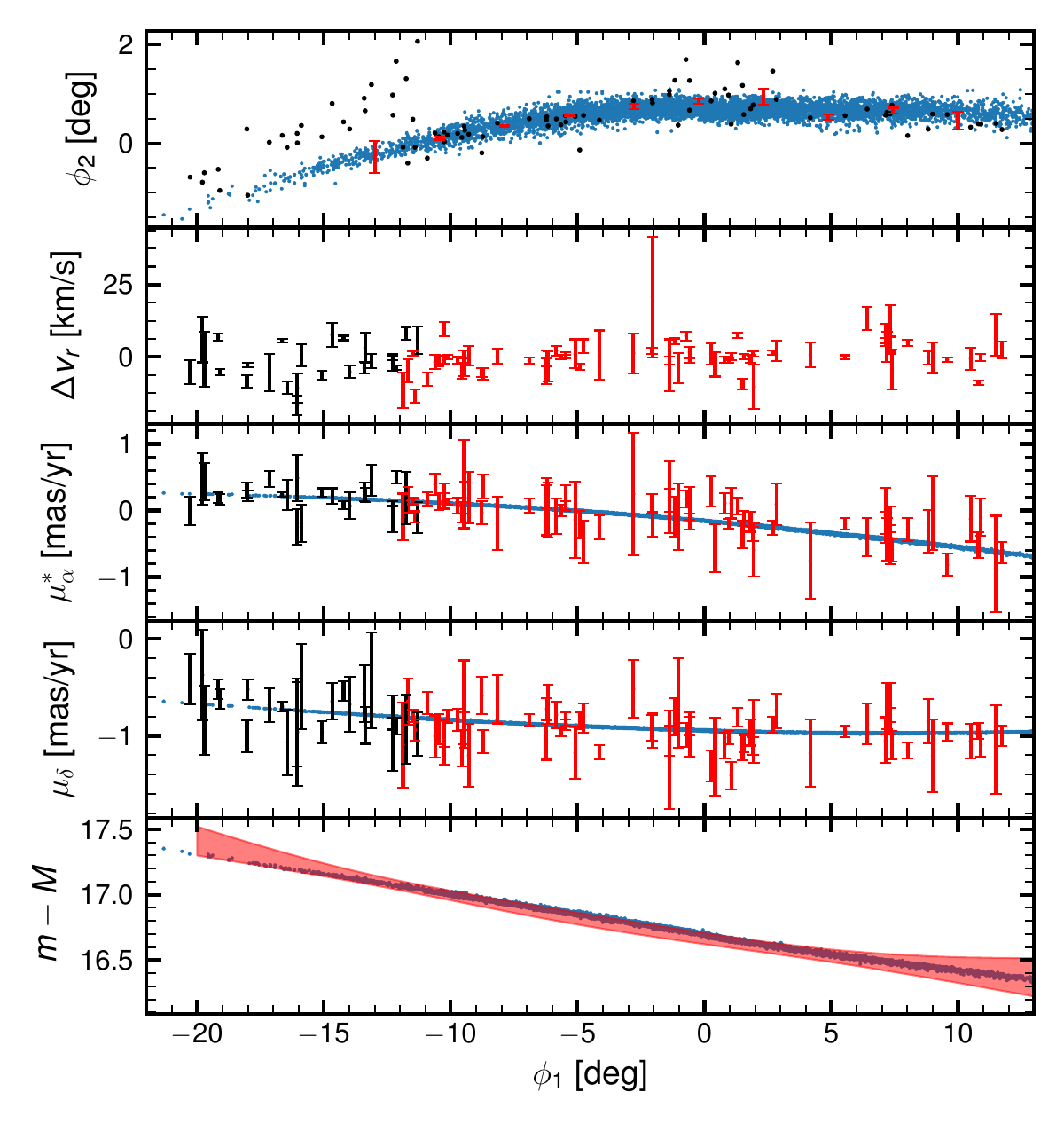}
    \caption{Best-fit stream model to the ATLAS stream. In each panel, the blue points show the best-fit stream model, the red-points show the data that were used in the fit, and the black points show data that was not used in the fit. \textit{Top panel} shows the stream on the sky. \textit{Second panel} shows the radial velocity difference between the observations and the model for clearer presentation, because the radial velocity spans a wide range. \textit{Third panel} and \textit{fourth panel} shows the proper motion in right ascension and declination respectively. \textit{Bottom panel} shows the distance modulus to the stream. The red shaded region shows the 1-$\sigma$ uncertainty on the distance modulus. Note that the continuation of the ATLAS stream model is a good match to most of the observed properties of the Aliqa Uma stream apart from the track on the sky (top panel).
    }
    \label{fig:atlas_sim}
\end{figure}

Figure \ref{fig:atlas_sim} shows the best-fit stream model compared to the data. In each panel we show mock observations of the simulated stream against the observations. For the radial velocity component, the difference between the observed data and the model ($\Delta v_r$) is shown for better presentation. The model fits the data along the ATLAS stream well. It also matches the observed properties of Aliqa Uma apart from the track on the sky, showing that these two streams are one and the same. 

This model highlights the peculiar features observed in the ATLAS stream discussed in Sections \ref{sec:spec} and \ref{sec:photometry}. First, the model does not capture the increased width or the wiggle in the stream track at $\phi_1 \sim 3\degr$. Furthermore, near the connection between ATLAS and Aliqa Uma ($\phi_1\sim-12\degr$) the observed radial velocity is more negative than the simulated velocity, supporting the interpretation in Figure \ref{fig:vdisp} that the radial velocity shows signs of a perturbation. Finally, this model passes through the possible continuation of AAU to $\phi_1 \sim 20\degr$ shown in Figure \ref{fig:gaia_map}.

We can also use the results of the MCMC to measure the orbital properties of the AAU stream. We find a pericenter of $13.3^{+0.1}_{-0.2}$ kpc, an apocenter of $41.0^{+0.4}_{-0.5}$ kpc, an eccentricity of $0.511\pm{0.001}$, and an orbital period of $0.62\pm{0.01}$ Gyr. The stream is on a prograde orbit with respect to the Milky Way disk. The present-day angular momentum of the progenitor has an orientation of ($\phi,\psi$)=($-11.2^{+0.4}_{\,\,\,-0.3},-24.3^{+0.2}_{\,\,\,-0.3})\degr$ where $\phi,\psi$ are the longitude and latitude as viewed from the Galactic center. As a consistency check, we also fit a plane to the best-fit stream particles in the observed range ($-20\degr < \phi_1 < 10\degr$) through the Galactic center and found a normal orientation of ($-5.2\degr,-24.9\degr$). This slight misalignment of the stream plane and its angular momentum is due to the effect of the LMC. We note that the orientation of AAU is broadly similar to the plane found in \cite{Shipp:2018} for ATLAS, who found ($-22.7\degr,-21.5\degr$) using photometric data from DES and to \cite{Pawlowski:2014} who found ($-21.9\degr,-24.8\degr)$ using the endpoints of the stream. Given this similar orientation, it is likely that ATLAS is still consistent with being a member of the vast plane of satellites \citep{Pawlowski:2014,Riley:2020}.

\section{Discussion}\label{sec:discuss}

\subsection{ATLAS and Aliqa Uma as One Stream}\label{sec:onestream}

\revise{The line-of-sight velocities and proper motions of the ATLAS and Aliqa Uma streams are seamlessly connected (Section \ref{sec:member}); the distance gradient observed in both streams are consistent with the one stream being slightly farther than the other one (Section \ref{sec:distance}); moreover, the metallicities and other elemental abundances are very similar between the two streams (Section \ref{sec:feh} and Section \ref{sec:abundance}). 
In order to quantify the similarity of the streams' kinematics, we use the best-fit stream model from Section \ref{sec:modeling} (see Fig. \ref{fig:atlas_sim}). We stress that this model was only fit to the ATLAS portion of the stream. For the stars associated with Aliqa Uma, we compute the difference in radial velocity and proper motions between this best-fit model and the observations. We fit the residuals with a Gaussian and find an offset of $-1.6\pm1.3 \kms$, $1.7\pm4.2 \kms$, and $1.5\pm5.1 \kms$ for the radial velocity, $\mu_\alpha^*$, and $\mu_\delta$ respectively. Note that to convert the proper motion residuals into a velocity, we have conservatively assumed a distance of $30$ kpc for Aliqa Uma. Thus, the kinematics of Aliqa Uma are consistent with it being part of the ATLAS stream. If these are two distinct streams, then they are on nearly identical orbits, whose kinematics differ at the level of $\sim 1 \kms$.}

\revise{Based on this evidence, we conclude that the ATLAS and Aliqa Uma streams are highly likely to be one stream or share one common origin. Although the possibility of two streams originated from two globular clusters from the same group infall cannot be ruled out completely, we argue that two globular clusters with identical metallcities and chemical abundances, and nearly identical orbits, are extremely unlikely. In particular, if the globular clusters were accreted with a dwarf galaxy, this dwarf would need a velocity dispersion on the order of $\sim 1 \kms$ to naturally explain the similarity of ATLAS's and Aliqa Uma's kinematics. Furthermore, in order to make two different streams almost connected but not have large overlap on the sky, the two globular clusters need to be disrupted at a particular time in which it was not too long time ago so that the two streams have no significant overlap, nor too recent so that there is a large gap between two streams; the chance of such a coincidence is extremely low.}

\subsection{Alignment of the AAU Stream}\label{sec:misalignment}

\begin{figure}
    \centering
    \includegraphics[width=0.49\textwidth]{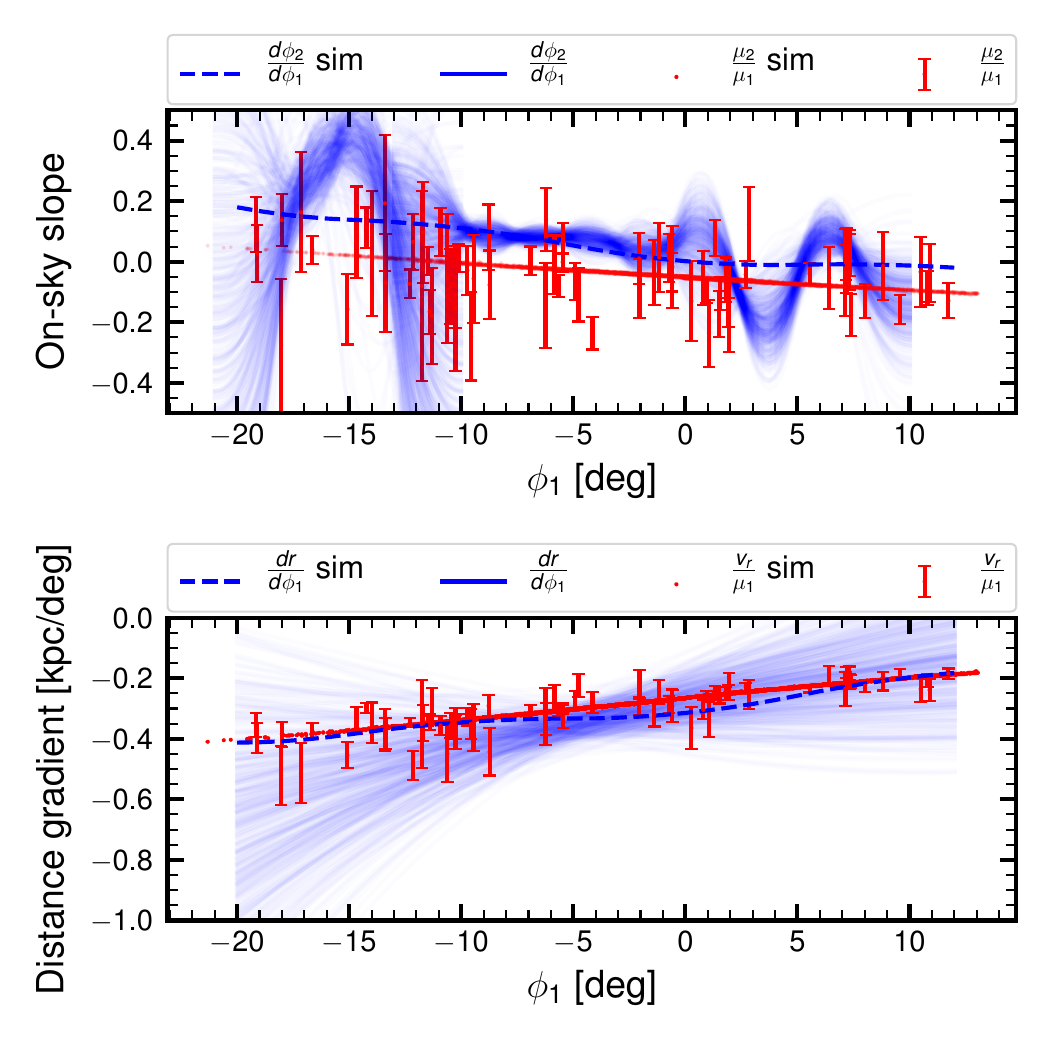}
    \caption{Alignment of the AAU stream's velocity and shape. \textit{Top panel} compares the stream velocity and shape on the sky which shows that the stream is increasingly misaligned for smaller $\phi_1$. The solid and dashed blue lines show the slope of the stream on the sky ($\frac{d\phi_2}{d\phi_1}$) in the data and best-fit simulation respectively. For the data we show 500 realizations of the slope drawn from the spline fit in Section \protect\ref{sec:DES_phot}. The red error bars and red points show the ratio of the reflex corrected proper motions ($\frac{\mu_2}{\mu_1}$) in the data \revise{(marginalized over distance and proper motion uncertainties)} and best-fit simulation respectively. \textit{Bottom panel} compares the stream velocity and shape along the line of sight which shows that the stream is broadly aligned in this direction. However, we note that there is a large uncertainty in the distance gradient. The solid and dashed blue lines show the distance gradient of the stream on the sky ($\frac{dr}{d\phi_1}$) in the data and best-fit simulation. For data, we show 500 realizations of the distance gradient drawn from the polynomial fit in Eq. \protect\ref{eq:distance} and its associated covariance. The red error bars and red points show the ratio of the reflex corrected radial velocity to the proper motion along the stream ($\frac{v_r}{\mu_1}$) in the data \revise{(marginalized over distance, proper motion, and radial velocity uncertainties)} and best-fit simulation. For the best-fit simulation, there is a slight misalignment near $\phi_1\sim0\degr$ due to the location of the progenitor. Note that in both panels we have only included stars with $g < 19$. 
    }
    \label{fig:atlas_align}
\end{figure}

Using the 6D view of the AAU stream from this paper, we can look at the alignment of the stream and whether the velocity is aligned with the shape of the stream. In particular, we follow the approach of \citet{Erkal2019} and \citet{deBoer2019} who showed that the alignment can be compared on the sky and along the line of sight. For the on-sky alignment, we compare the slope of the stream on the sky ($\frac{d\phi_2}{d\phi_1}$) \revise{using stream track derived in Section \ref{sec:DES_phot}} with the ratio of reflex corrected proper motions ($\frac{\mu_2}{\mu_1}$) \revise{from individual spectroscopic members}. We stress that $\mu_1$ does not contain the typical $\cos(\phi_2)$ term. We make this comparison in the top panel of Figure \ref{fig:atlas_align} \revise{which shows that the slope of the stream track (solid blue lines) is misaligned with the ratio of the on-sky tangential velocities (red points with error bars)}. For the ATLAS portion of the stream ($\phi_1 > \sim -13\degr$), this misalignment matches the misalignment in the simulation on average, \revise{shown as the dashed blue line and small red points.
In models without the LMC, the stream shape and velocity slope are aligned (i.e. the blue line and red points lie on top of each other. The offset/misalignment is due to the effect of the LMC.}

In order to compare the alignment along the line of sight, in the bottom panel of Figure \ref{fig:atlas_align} we show the distance gradient of the stream ($\frac{dr}{d\phi_1}$)(in blue) with the ratio of the Solar reflex corrected velocity and proper motion ($\frac{v_r}{\mu_1}$) (in red). This shows that the velocity is aligned with the stream along the line of sight, as is expected from the simulation. However, since the uncertainties are large, improving the distance gradient will make this comparison more meaningful. We note that the misalignment in the simulation at $\phi_1 \sim 0\degr$ is due to the progenitor.

\subsection{Perturbation by baryonic substructures}\label{sec:perturb}

In order to check whether the perturbations in AAU could be due to baryonic substructure in the Milky Way, we consider a variety of perturbers which can affect streams. In particular, we consider the effect of the bar \citep[e.g.][]{Hattori:2016,Price-Whelan:2016,Erkal:2017,Pearson2017}, spiral arms \citep{Banik:2019}, giant molecular clouds \citep[GMCs, ][]{Amorisco2016}, classical satellites, and globular clusters. Interestingly, while a number of these mechanisms can create subtle features in the stream,
we find that of the mechanisms considered, only the Sagittarius dwarf is capable of creating the kink feature. 

\subsubsection{Milky Way bar} \label{sec:bar}

For the bar we consider the analytic bar potential from \cite{LongMurali:1992}. Following \cite{Hattori:2016,Erkal:2017} we use a semi-major axis of $a=3$ kpc and a semi-minor axis of $b=1$ kpc for the bar. For the mass, we use the recent results of \cite{Portail:2017} and take a bar mass of $10^{10} \mathrm{M}_\odot$. For the pattern speed, we use $\Omega=41\pm3\,\kms\,\mathrm{kpc}^{-1}$ from \cite{Sanders:2019} which is consistent with other recent measurements \citep[e.g.][]{Portail:2017,Bovy:2019}. We take the bar's present-day orientation to be $30\degr$ \citep{Wegg:2015}. When including the bar, we set the bulge mass to zero. 

In order to account for the uncertainty in AAU's orbit, we sample the MCMC chains from Section \ref{sec:modeling} 100 times. For each of these samples, we also sample the bar's pattern speed from its observed value and uncertainty. Since the bar slightly changes the mass distribution of the Milky Way potential, we compare these streams with those disrupted in the presence of a rapidly rotating bar ($\Omega = 1000\,\kms\,\mathrm{kpc}^{-1}$). 

For each of the 100 realizations, we compute the change in the stream track measured at the $\phi_1$ locations of the nodes from the fit in Figure \ref{fig:posterior}. The maximum change amongst all realizations is $0.1\degr$ and the median of the maximum change for each realization is $0.03\degr$. This shows that the bar is not capable of creating the kink between ATLAS and Aliqa Uma. Similarly, we compare the stream density in 1 degree bins and find that the median of the maximum change in the density is $\sim25\%$. Thus, while the bar should not have a significant effect on the stream track of AAU, it can create modest density variations.

\subsubsection{Giant Molecular Clouds (GMCs)}

For the GMCs we take a similar approach to \cite{Banik:2019}. In particular, we take the catalog of observed GMCs from \cite{Miville-Deschnes:2017}. Since AAU has a pericenter of $\sim 13$ kpc, we only consider the GMCs with galactocentric radii beyond 10 kpc. We only consider GMCs with mass greater than $10^5 \mathrm{M}_\odot$ since perturbers below this mass will not create significant features in the stream \citep{Erkal:2016,Bovy:2017}. As in \cite{Banik:2019} we consider the GMC population within the same quadrant as the Sun which is the most complete. However, instead of replicating this quadrant, for each GMC in this patch we create 4 copies by randomly sampling its azimuthal angle. This gives 624 GMCs beyond 10 kpc with a mass larger than $10^5 \mathrm{M}_\odot$. We model each GMC as a Plummer sphere with the observed mass and a scale radius which is one-third that of the observed size. This reduced size means that 90\% of each GMC's mass is within the observed size \citep{Banik:2019}. Each GMC is then placed on a circular orbit in the plane of the disk. The influence of all GMCs is included during the rewinding procedure and subsequent stream generation.

As with the bar in Section \ref{sec:bar}, we consider the same 100 realizations of the AAU stream in order to account for the variation in the stream orbit. The addition of these GMCs slightly changes the mass distribution of the Milky Way potential so we once again consider a rapidly rotating population of GMCs as our fiducial setup to account for the smooth change in the potential. To do this, we keep the GMCs on their original circular orbits but increase the angular velocity by a factor of 100. As with the bar, we compare the change in the stream track and the stream density. For the stream track, we get a maximum difference of $0.04\degr$ and a median of the maximum change for each realization of $0.008\degr$. Thus the present day distribution of GMCs do not appear to be capable of creating the kink. This is due to a combination of the modest mass of the GMCs as well as the assumption that the GMCs are confined to the Milky Way plane while AAU is on a highly inclined orbit. As a result, there will always be a significant relative velocity between AAU and the GMCs at closest approach which will limit the size of the perturbation \citep[e.g.][]{Erkal:2015a}. The median of the maximum density change is $\sim20\%$, indicating that GMCs can also make modest density features in the stream.

\subsubsection{Spiral arms} \label{sec:spiral_arms}

In order to assess the impact of spiral arms, we follow largely the same procedure as \cite{Banik:2019}. Namely, we use the analytical spiral arm potential from \cite{Cox:2002} and implement it following a sinusoidal density distribution. As in \cite{Monari:2016}, we use tightly wound spirals with a constant pitch angle of $9.9\degr$ and fix their amplitude such that the maximum force from the spirals at a distance of 8 kpc from the Galactic center is 1\% of the disk force at that distance. This amplitude is determined using spirals arms with scale lengths and heights of 3 kpc and 0.3 kpc respectively, as used by \cite{Banik:2019}. We randomly sample the pattern speed 100 times from a Gaussian with $\Omega_{\rm spiral}=22\pm2.5\,\kms\,\mathrm{kpc}^{-1}$. As with the Milky Way bar in Section \ref{sec:bar}, we consider a fiducial setup with a pattern speed of $\Omega_{\rm spiral}=1000\,\kms\,\mathrm{kpc}^{-1}$ to account for any smooth change to the potential due to the spirals. For the stream track, we find a maximum change of $0.02\degr$ and for the density there is a median maximum change of 7\%. This shows that spiral arms cannot significantly affect the AAU stream.

\subsubsection{Classical satellites} \label{sec:dwarfs}

In order to assess the impact of the 10 classical satellites (excluding the LMC), we include each satellite as an additional perturber. Motivated by the results of \cite{Law:2010}, each satellite is modeled as a $10^9 M_\odot$ Plummer sphere with a scale radius of $1$ kpc. This is not meant to perfectly represent each satellite, but rather to check whether they can create a feature qualitatively like the kink. We note this neglects the effect of the tidal debris from the dwarf on AAU \citep{Bovy:2016b} which may be important in the event of a close flyby. For the proper motions, we use the results of \cite{GaiaDR2Kinematics} except for Leo I, Leo II, and the SMC for which we use proper motions from \cite{Sohn:2013,Piatek:2016,kallivayalil13}, respectively. The other properties come from \cite{McConnachie2012} and references therein. For simplicity, we do not consider ultra-faint dwarfs.

We use the same 100 realizations of AAU's orbit from Section \ref{sec:bar}. For each realization, we sample the observed properties of each dwarf (i.e. distance, radial velocity, and proper motions). The effect of the dwarf on the progenitor, Milky Way, and LMC is included during the rewinding procedure and on the stream during the disruption. Note that we consider the effect of each of the 10 dwarfs separately so this results in 1000 stream disruptions. For each stream, we compute the change in the stream track and the stream density. We find that only Sagittarius can have a large effect on the stream track with a maximum track deviation of $\sim 1\degr$ while the other dwarfs have a maximum deviation of $0.06\degr$. Interestingly, 6 of these realizations of Sagittarius produce kink-like features in AAU, although not at the observed location of $\phi_1\sim-12\degr$.

In order to study the effect of Sagittarius more closely, we take the phase-space coordinates (i.e. proper motions, distance, and radial velocity) of one of the original realizations which produces a kink and resample about these values 1000 times with 10\% of the observed uncertainties. We then make mock observations of these in each observable (e.g. as in Figure \ref{fig:atlas_sim}) and select those with a kink at $\phi_1 \sim -12\degr$ based on visual inspection.

\begin{figure}
    \centering
    \includegraphics[width=0.49\textwidth]{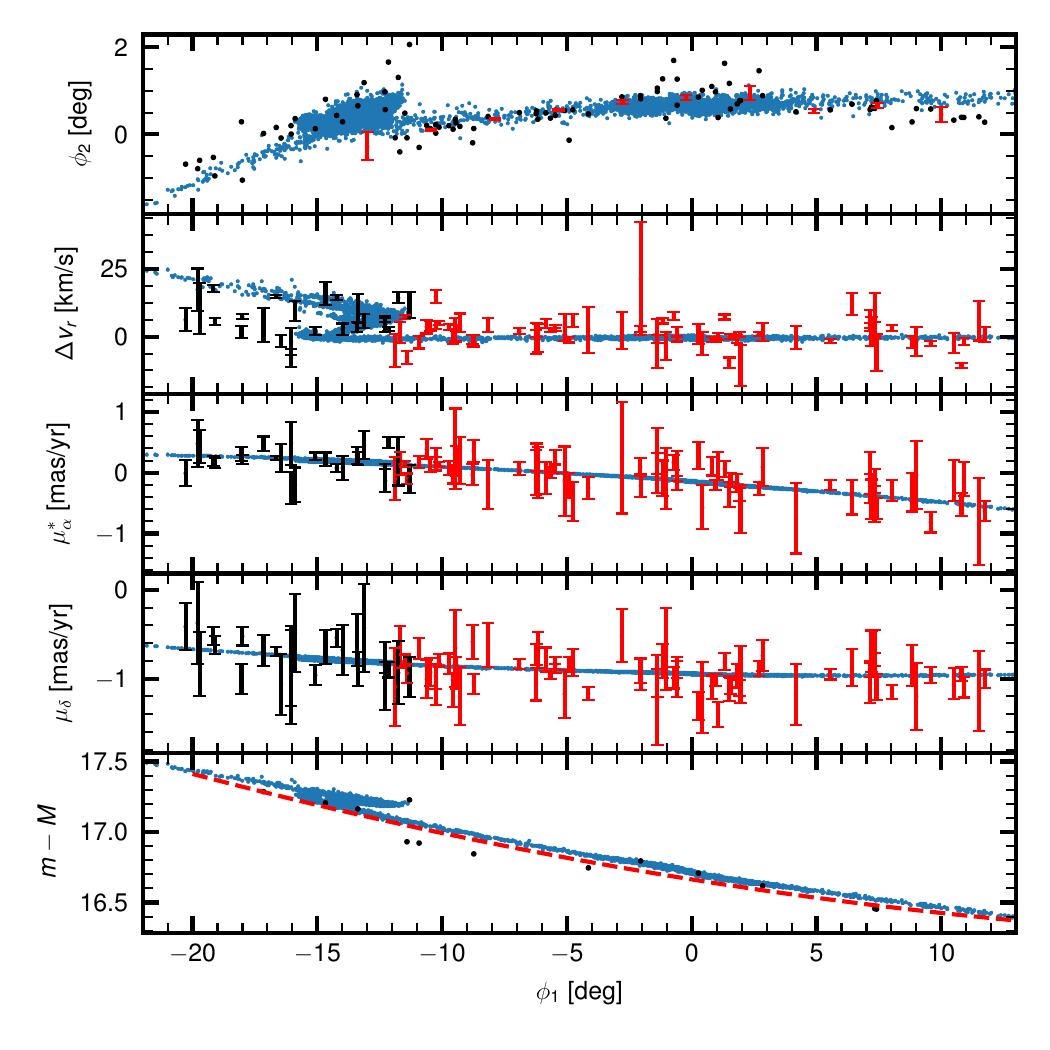}
    \caption{Example of perturbation from the Sagittarius dwarf on the ATLAS stream. This realization was chosen to have a kink at $\phi_1 \sim -12\degr$ (see text for details). The panels show the same mock observables as in Figure \protect\ref{fig:atlas_sim} apart from the radial velocity where we show the difference from a quadratic function fit to the simulated stream between $-10\degr < \phi_1 < 13\degr$. In addition, we show the distance modulus of individual BHBs from Figure \protect\ref{fig:dmgrad} in the bottom panel.  Interestingly, this perturbation also produces a kink in the radial velocity and distance modulus similar to the observations although we note that the radial velocity in the model to the left of $\phi_1 < -15\degr$ does not match the observed trend.}
    \label{fig:atlas_sgr}
\end{figure}

\begin{figure*}
\centering
\begin{interactive}{animation}{atlas_sgr.mp4}
\includegraphics[width=0.79\textwidth]{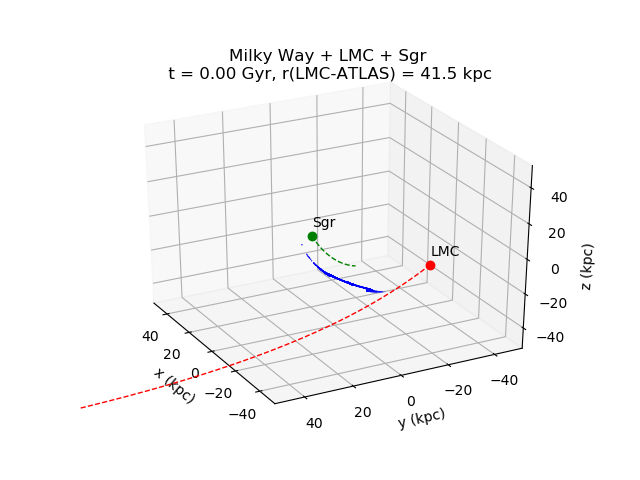}
\end{interactive}
\caption{A movie showing a perturbation from the Sagittarius dwarf on the AAU stream which can qualitatively reproduce the kink between the ATLAS stream and the Aliqa Uma stream. The present-day observables of this model are shown in Fig. \ref{fig:atlas_sgr}. In this movie, the AAU stream is shown in blue, the dashed-green (red) line shows the past orbit of the Sagittarius dwarf (LMC), and the green (red) circle shows the Sagittarius dwarf's (LMC's) present day location. This figure is available as an animation in the HTML version of the final article. The animation can also be viewed at \url{https://youtu.be/GjZJYEQQZXU}.}
\label{fig:movie}
\end{figure*}

Figure \ref{fig:atlas_sgr} shows the mock observations of present day for one of these realizations that qualitatively matches the observed properties of AAU with a kink in the stream track, a $\sim 10$ km/s change in the radial velocity, and a kink in the distance modulus all at $\phi_1 \sim -12\degr$. We also note that the model does not match the radial velocity to the left of $\phi_1 \sim -12\degr$.
A movie of this simulation is presented in Figure \ref{fig:movie}.
This kink is the result of a close approach between Sagittarius and AAU $\sim0.51$ Gyr ago at a distance of $\sim0.9$ kpc with a relative velocity of $\sim 400$ km/s. The closest approach changes the orbital period of particles in the stream and creates a gap with particles piling up at the edge of the gap \citep[e.g.][]{Erkal:2015a}. One of these pile-ups occurs at $\phi_1 \sim-13\degr$ and creates the kink and overdensity. The other pile-up is located at $\phi_1 \sim25\degr$ which is beyond the currently observed range of AAU. 


We note that given the current uncertainties on the present-day phase-space position of Sagittarius, we cannot definitively determine whether or not it has interacted with AAU in the past. In order to explore this, we computed where the past orbit of Sagittarius (using the realizations above) passed through the stream plane of AAU given the uncertainty in the proper motion, radial velocity, and distance of Sagittarius. These crossings occur $0.4\pm0.1$ Gyr ago with an uncertainty of $3.0$ kpc in where they cross the AAU stream plane. This is mostly driven by the distance uncertainty; improving the distance errors by a factor of 2 lowers this uncertainty to 1.5 kpc. Interestingly, this uncertainty in crossing the AAU stream plane does not seem to be heavily affected by the uncertainty in the Milky Way potential. We explored this by also sampling from the posterior samples from \cite{McMillan:2017} and found the same uncertainty of $3.0$ kpc. Thus, improved measurements of the phase-space location of Sagittarius will help us determine whether it created the kink in AAU.

Finally, we note that \cite{deBoer2019} have also shown that the Sagittarius dwarf could have perturbed the GD-1 stream \citep{Grillmair:2006_GD1}. If it can be shown that Sagittarius perturbed both GD-1 and AAU, this would place very tight constraints on the orbit of Sagittarius as well as the potential of the Milky Way.

\subsubsection{Globular clusters} \label{sec:gc_perturbations}

Similar to the classical dwarf galaxies in Section \ref{sec:dwarfs}, we also consider the population of globular clusters in the Milky Way as potential stream perturbers. For this we use the globular cluster catalog of \cite{vasiliev_GCs} which gives the 6D phase-space positions of 147 globular clusters. For each of the 100 realizations of AAU's orbit from Section \ref{sec:bar}, we sample the observed properties of each globular cluster and include the cluster during the rewinding and stream disruption process. As with the dwarfs in Section \ref{sec:dwarfs}, we include the globular clusters one at a time so this results in $14700$ stream disruptions. To be conservative, we model each cluster as a Plummer sphere with a mass of $10^6 M_\odot$ and a scale radius of $10$ pc. 

For each simulation, we measure the simulated stream track and density, as well as how close the cluster comes to each stream particle. Four globular clusters have a median closest approach within 2 kpc: Pal 12 (1.9 kpc), NGC 5904 (1.5 kpc), NGC 6229 (1.4 kpc),  and NGC 7492 (0.6 kpc).

Furthermore, we find that 16 globular clusters have closest approaches (amongst their 100 realizations) within 100 pc of the stream. For most of these globular clusters, only 1 out of 100 of the realizations pass within 100 pc, indicating that this is due to significant uncertainty in the past trajectory. However, NGC 7492 and NGC 6229 stand out, having a 17\% and 7\% chance of passing within 100 pc of the stream respectively.

In terms of the stream track, 8 globular clusters produce deviations which are larger than $0.1\degr$ with a maximum deviation of $0.24\degr$. Of these, one (NGC 7492) produces a feature like a kink in the stream track with a deviation of $0.19\degr$. We show this in the top panel of Figure \ref{fig:atlas_NGC7492} in Appendix \ref{sec:gc}, while the other panels show other perturbations from NGC 7492. Interestingly, some of these realizations also exhibit a broadening of the stream track similar to the one observed at $\phi_1 \sim 3\degr$ (see Figure \ref{fig:modelcomparison}). 
We note, however, that in the January 2020 version of the \citet{Baumgardt:2019} catalogue of fundamental parameters of Galactic globular clusters\footnote{https://people.smp.uq.edu.au/HolgerBaumgardt/globular/}, the mass of NGC 7492 is listed as $2.8 \pm 0.8 \times10^4 M_\odot$ which is significantly smaller than the mass we have assumed. Thus, while globular clusters may be able to create a subtle feature in AAU, like the broadening, they cannot create the large kink at $\phi_1\sim-12\degr$.

\subsubsection{Progenitor}

Using the best-fit stream from Section \ref{sec:modeling}, we can assess whether any of the features in the data are consistent with the progenitor. At the location of the progenitor, the stream will connect on at the inner and outer Lagrange points \citep[e.g.][]{Combes:1999}, which can cause a visible kink in the stream \citep[e.g. Pal 5, ][]{Odenkirchen:2001} depending on the orientation of the stream relative to the observer. For AAU, the angle between the line of sight and the radial direction from the Galactic center is $49.0\degr$ at $\phi_1 = 0\degr$ suggesting that if a progenitor was present, we would be able to see the stream connecting onto the progenitor which would appear as a wiggle near the progenitor. In order to explore this, we re-simulate the best-fit AAU model from Section \ref{sec:modeling} with progenitor masses of $2,20,200\times10^4 \mathrm{M}_\odot$ and force the progenitor mass to remain constant throughout the simulation. These give significant wiggles in the stream track with sizes of $0.26\degr,0.56\degr,$ and $1.2\degr$ respectively. 
In order to match the $\sim 2\degr$ size of the kink between ATLAS and Aliqa Uma, we would need a present-day progenitor mass of $\sim 8\times10^6 \mathrm{M}_\odot$, over 3${\times}$ more massive than the most massive known globular cluster and thus certainly ruled out \citep{Harris2010}. 

\subsection{Connection to other globular clusters}

In order to assess the relation between the AAU stream and globular clusters in the Milky Way, we compute the actions of our best-fit stream and each globular cluster. For each globular cluster, we sample the observed proper motions, distances, and radial velocities 100 times given their uncertainties to get the spread in actions. For the observed properties we use the globular cluster catalog from \cite{vasiliev_GCs}, which contains 147 globular clusters. Note that we have replaced the distance to Palomar 5 with an updated distance of $20.6\pm0.2$ kpc from \cite{Price-Whelan:2019}. We compute the actions using \code{AGAMA} \citep{Vasiliev_Agama}. 

\begin{figure*}
    \centering
    \includegraphics[width=0.98\textwidth]{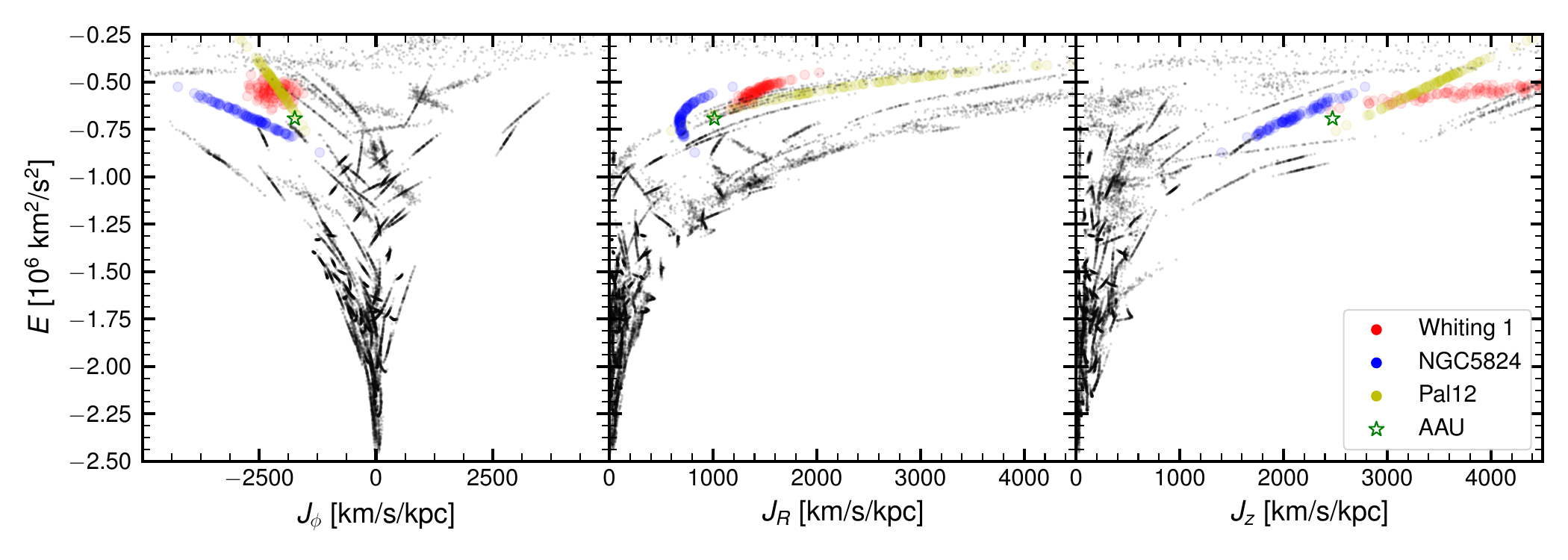}
    \caption{Actions for 147 Milky Way globular clusters and the best-fit to the ATLAS stream. For each globular cluster, we have sampled their actions 100 times given the uncertainties on their present day proper motions, distances, and radial velocities using the catalog from \cite{vasiliev_GCs}. The green star shows the actions for our best-fit orbit for the AAU stream. The red, blue, and yellow points show the actions of the globular clusters closest to AAU in action space, Whiting 1, NGC 5824, and Pal 12, respectively. Interestingly, these are all associated with the Sagittarius dwarf \protect\citep[e.g.][]{Irwin:1999,Bellazzini:2003,Carraro:2007,Massari:2019}, suggesting that the progenitor of the AAU stream may have been accreted with the Sagittarius dwarf. The black points show the actions of the remaining 144 globular clusters.}
    \label{fig:atlas_action}
\end{figure*}

In Figure \ref{fig:atlas_action} we show these actions along with that of the AAU stream. We compute the distance between AAU and each globular cluster in action space using the combined action modulus,

\begin{equation}
\Delta J =  \sqrt{\Delta J_\phi^2+\Delta J_R^2 + \Delta J_z^2}
\end{equation}
We have highlighted the three globular clusters closest in action space: Whiting 1, NGC 5824, and Pal 12. Interestingly, these have previously been associated with the Sagittarius dwarf \citep[e.g.][]{Irwin:1999,Bellazzini:2003,Carraro:2007,Massari:2019}. Furthermore, the eccentricity and apocenter of AAU stream is very similar to the Sagittarius GCs discussed in \cite{Kruijssen2020}, suggesting that the progenitor of the AAU stream may have been originally been bound to the Sagittarius dwarf. In further support of this, we note that the mean metallicity of AAU is similar to that of one of the GCs associated with Sagittarius, Terzan 8 \citep[e.g.][]{Massari:2019}, which has a metallicity of $\feh \sim -2.27$ \citep{Carretta:2014}.

\subsection{Complex stream morphologies}

Recent works have shown that almost every stream studied in detail has signs of a significant perturbation. Pal 5 shows clear gaps which are inconsistent with evolution in a smooth, time-independent potential \citep[e.g.][]{Erkal:2017,Bonaca2020}. GD-1 has a spur of stars that run parallel to the stream and a blob of co-moving stars below the stream, as well as wiggles and density variations \citep[e.g.][]{deBoer2018,Price-Whelan2018,Malhan2019,deBoer2019}. The Ophiuchus stream also exhibits a spur-like feature parallel to the main track \citep{Caldwell:2020}. This appears to support the models of \cite{Carlberg:2020}, which predict that globular cluster streams have a rich morphology due to their initial disruption in their host dwarf galaxy before being accreted into the Milky Way. 

Similarly, streams from dwarf galaxies also show rich structures.  The Sagittarius stream exhibits a prominent bifurcation \citep{Belokurov2006b} and the Jhelum stream appears to have multiple components \citep{Bonaca:2019,Shipp2019}. In addition, the Orphan stream has a substantial velocity perpendicular to the stream \citep{Fardal2019,Koposov2019} due to the perturbation from the LMC \citep{Erkal2019}. Similarly, many of the streams discovered in DES exhibit substantial misalignment between the stream track and the on-sky velocity, likely due to the LMC \citep{Shipp2019}, including the AAU stream, as we discussed in Section \ref{sec:misalignment}.

\subsection{Palca stream in the Aliqa Uma stream field}\label{sec:palca}

\begin{figure*}
    \centering
    \includegraphics[width=0.9\textwidth]{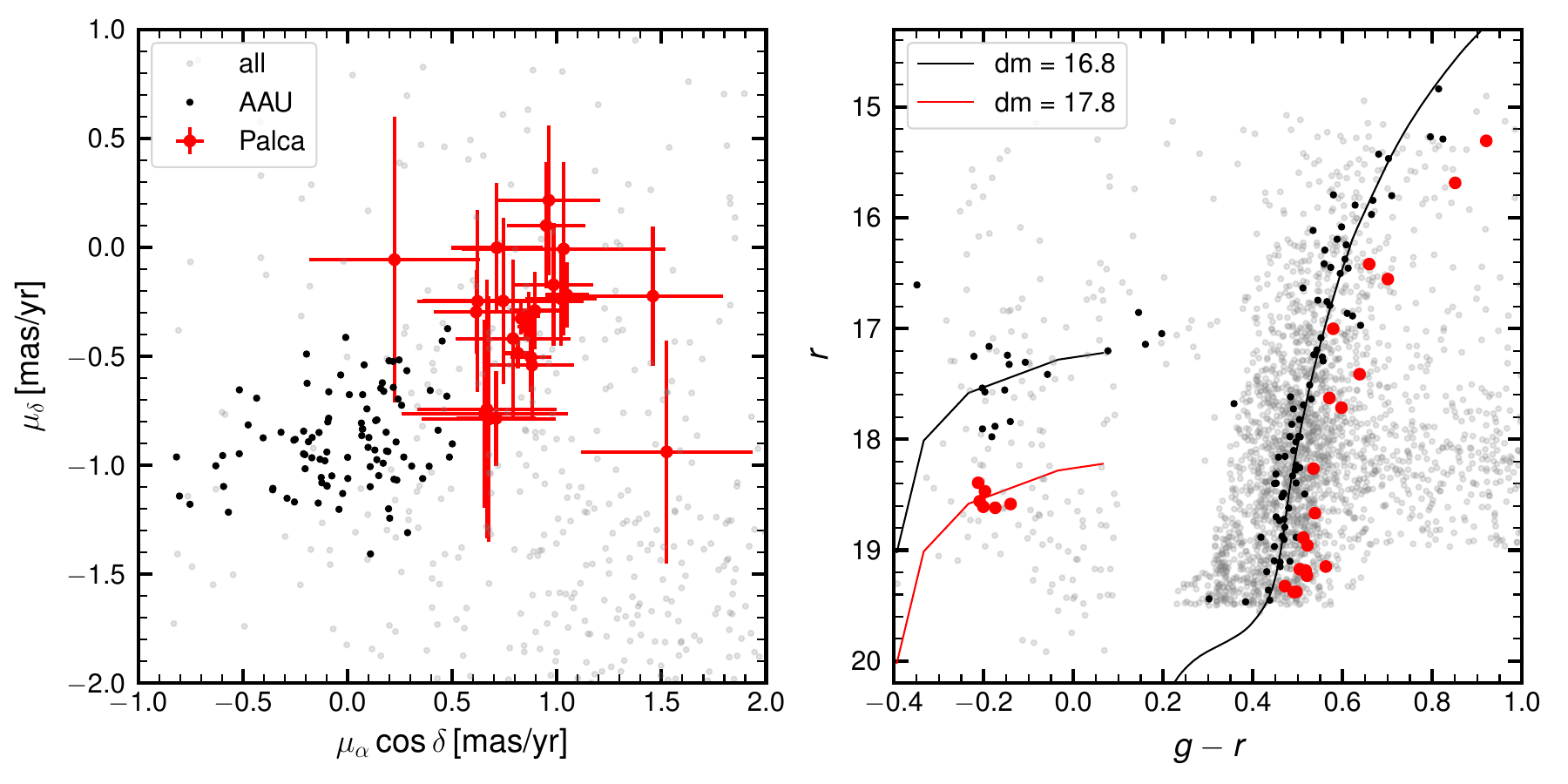}
    \caption{Proper motions (left) and CMD (right) of the other structure discussed in Figure \ref{fig:vhel_dist}. Stars with $80<\vhel < 130\,\kms$ are selected and shown as the red circles with error bars. These stars are also clustered in proper motion space with a distance modulus of 17.8, and are likely to be member stars of the Palca Stream. The grey dots show all observed stars, while black dots represent the spectroscopic members in the AAU stream.
    }
    \label{fig:palca}
\end{figure*}

As discussed at the beginning of this paper, in Figure \ref{fig:vhel_dist}, when selecting high priority candidate members in AAU, we also see substructure in velocity around $\vhel\sim100\,\kms$, especially in the fields of the Aliqa Uma stream. These stars are also clustered in proper motion space around 
$$\mu_\alpha cos\delta = 0. 85\,\masyr$$
$$\mu_\delta = -0.37\,\masyr$$
The proper motion is very close to the AAU stream and therefore some were selected as high priority candidates. Figure \ref{fig:palca} shows the stars with the following selection criteria
$$80<\vhel < 130\,\kms$$
$$ |\mu_\alpha cos\delta - 0.85| < \mathrm{max}(0.3, 2 \sigma_{\mu,\alpha})$$
$$ |\mu_\delta + 0.37| < \mathrm{max}(0.3, 2 \sigma_{\mu,\delta})$$
and 
$$-20\degr < \phi_1 <-10\degr$$
We found a very clear stellar association at a distance modulus of $m-M\sim 17.8$ in the CMD (right panel of the Figure), further confirming that this is a real structure rather than just a random clustering in line-of-sight velocities.

Given the distance and the location on the sky, this structure is very likely to be the Palca stream, which was also discovered in DES \citep{Shipp:2018}. Recent studies by \citet{Chang2020} show that Palca is possibly the extension of Cetus Polar stream found in SDSS \citep{Newberg2009,Koposov:2012}. The kinematic data will help confirm or refute this connection. If it is indeed one stream, this long stream with 6D information will be another critical tool for constraining the Milky Way potential.

We found a total of 25 Palca member stars using the selection criteria defined above (listed in Table \ref{table:palca_camembers}, which gives a velocity dispersion of $\sigma_v = 9.5\pm1.8\,\kms$ and a systemic velocity of $\vhel = 98\pm2\,\kms$ at $(\alpha, \delta) = (34\degr, -34\degr)$. Based on the large velocity dispersion, the progenitor is very likely to be a dwarf galaxy, which matches with the large stream width observed on the sky. 
We derived the metallicity of the 11 brightest RGB members of Palca assuming a distance modulus of $m-M = 17.8$. These stars have metallicities spanning from \feh = $-$1.5 to \feh = $-$2.2, with a mean metallicity of $\feh = -2.0$. However, we were not able to resolve a metallicity dispersion ($\sigmafeh < 0.16$ dex at 95\% confidence). The low metallicity dispersion is likely due to a combination of small sample size and the faintness (and therefore large metallicity uncertainty) of the RGB stars. Although \SSSSS did not specifically target the Palca stream, many \SSSSS fields overlapped with it due to its large width on the sky. We will leave a more thorough analysis of Palca for a future \SSSSS paper.

\input{table_palca_specmem.tex}

\section{Summary}
\label{sec:summary}

We present the first spectroscopic measurements on the ATLAS stream and the Aliqa Uma stream from \SSSSS observations, with a total of 96 spectroscopic member stars identified in these two streams (Figure \ref{fig:cmd} and \ref{fig:specmem}). In combining our spectroscopy with the photometry from DES DR1 and PS1 DR1, and astrometry from \gaia DR2, we conclude that the two streams are \revise{extremely likely to be} one stream, despite the discontinuity in the on-sky morphology, \revise{although scenarios in which two streams originated from two globular clusters from one group infall cannot be completely ruled out}. We refer to this entire stream as the ATLAS-Aliqa Uma stream, or the AAU stream. We summarize our main findings here:

\begin{itemize}
  \item We confirm that in radial velocity, proper motion and heliocentric distance (see Figures \ref{fig:specsel}, \ref{fig:specmem} and \ref{fig:dmgrad}) the two streams are seamlessly connected to each other, with a $\sim1$\degr\ shift in the stream track on the sky at the connection point at $\phi_1\sim -12 \degr$; a feature we call a ``kink". The physical size of the kink feature is $\sim$0.5 kpc. 
  
 \item In addition to the ``kink", we notice a significantly larger stream width on the sky around $\phi_1\sim0\degr$ in the spectroscopic sample (Figure \ref{fig:specsel} and \ref{fig:specmem}). We call this feature ``broadening". This feature is well detected in a deep photometric map of the stream based on DES DR1 (without spectroscopic or proper motion information). 
 \revise{The modeling of the feature reveals two (surface) density gaps at $\phi_1\sim-2\degr$ and $\phi_1\sim+3\degr$ (Figure \ref{fig:modelcomparison}, \ref{fig:posterior}), in which the surface brightness of the stream drops by about a factor of two while the stream width gets larger. The resulting linear density of the stream members therefore is roughly constant in this ``broadening" area.}
 This feature is also accompanied by a detectable shift in the stream track (or referred to as ``wiggle") by 0.2\degr. The constant linear density combined with the shift in the stream track strongly supports a perturbation hypothesis as opposed to density variation caused by the epicyclic motion of the stripped stars \citep{Ibata2020}. 

  \item We find that the line-of-sight velocity dispersion varies along the stream. In the Aliqa Uma part (including the ``kink"), the velocity dispersion is as large as $\sim6\,\kms$, while in the ATLAS part of stream the dispersion is around $\sim 2\,\kms$. Furthermore, we also see an indication of the velocity gradient along $\phi_2$ at the "kink", where the line-of-sight velocities show a difference of $>20\,\kms$ at $\phi_1\sim-11\degr$ (Figure \ref{fig:vdisp}). This suggests that the Aliqa Uma component was heavily perturbed in the past, confirming the picture painted based on the discontinuity of the stream on the sky. 
  
  \item In addition to finding continuity between ATLAS and Aliqa Uma in kinematic space, we observe that they are indistinguishable in metallicity and chemical abundance patterns, further supporting the hypothesis that they are one stream. The mean metallicity of the stream is at $\feh = -2.2$, with an unresolved metallicity dispersion ($<0.07$ dex at 95\% confidence level). The low metallicity dispersion together with the narrow stream width and low velocity dispersion confirm the hypothesis that the progenitor of the stream was likely a globular cluster.
  
 \item In the list of high probable member stars identified with help of \gaia and DES we notice a possible extension of the Aliqa Uma stream that protrudes out of the stream track around $\phi_1\sim-10\degr$ and $\phi_2\sim+2\degr$ (Figure \ref{fig:gaia_map}). We call that feature a ``spur" as its shape is broadly similar to the feature seen in the GD-1 stream \citep{Price-Whelan2018}. As the \SSSSS observations did not cover this feature, further spectroscopic observations in this area are needed to confirm or disprove its existence. If this spur feature is real, it extends from the ATLAS stream by $\sim2\degr$ on the sky, or $\sim0.9$\,kpc  which is about a factor of 6 times larger than the separation between the spur and the main stream for GD-1 \citep{Price-Whelan2018}. 
  
  \item By mapping the probable member stars with proper motion from \gaia and photometry from DES DR1 and PS1 DR1, we find that the entire stream covers at least 40 degrees on the sky (Figure \ref{fig:gaia_map}). As the stream also spans from 20 kpc to 30 kpc in heliocentric distance (Figure \ref{fig:dmgrad}), the total visible portion of the stream is more than 20 kpc long. 
  
  \item Using the stream track and spectroscopic sample, we fit a dynamical model to the ATLAS component of the stream in the presence of the LMC and determined that the orbit of the AAU stream has a pericenter of $13.3^{+0.1}_{-0.2}$ kpc, an apocenter of $41.0^{+0.4}_{-0.5}$ kpc, an eccentricity of $0.511\pm{0.001}$, and an orbital period of $0.62\pm{0.01}$ Gyr. \revise{We further confirm that the kinematics of Aliqa Uma are consistent with the best fit model, and these two streams have nearly identical orbits, further confirming they are extremely likely to be one stream.} Using these orbit fits, we also compared the actions of the AAU stream with the Milky Way globular clusters and found that the stream has actions similar to globular clusters that were accreted with the Sagittarius dwarf (Whiting 1, NGC 5824, Pal 12). 
  
 \item We examine a wide range of baryonic effects on the AAU stream: the Milky Way bar, spiral arms, giant molecular clouds, globular clusters, and dwarf galaxies. Of these, we find that only a nearby passage with the Sagittarius dwarf can create features similar to the observed ``kink" between ATLAS and Aliqa Uma. In order to confirm this, a more detailed analysis is needed to fit the perturbed models of the AAU stream to the data and constrain the perturbation \citep[e.g.][]{Erkal:2015b}. We also find that the globular cluster NGC 7492 likely has a close passage with AAU and may be able to create features like the ``broadening". 
 
 \item In addition to the AAU stream, we found another group of stars in the observed fields at a heliocentric velocity of $\sim100\,\kms$ and a distance of $\sim35$\,kpc. This structure is unconnected to the AAU stream, and is very likely to be associated with the Palca stream (Figure \ref{fig:vhel_dist}, \ref{fig:palca}), another stream found in DES and possibly a southern extension of the Cetus Polar Stream.
\end{itemize}
 
We want to highlight that the ATLAS and Aliqa Uma streams are the second pair of streams that have been found to be a single, gravitationally perturbed stream. The first example of such a case was the Orphan/Chenab pair found in \citet{Koposov2019}. This significant result suggests that 1) many streams that are currently thought to be distinct could in fact have the same progenitor; 2) perturbations at small (for AAU) and large scales (for Orphan/Chenab) play a critical role in the  evolution of stellar streams.
 
The detection of the ``kink" and ``broadening" features show the power of spectroscopy as part of density variation studies for distant streams. Unlike the GD-1 stream, at a heliocentric distance of  $7-10$\,kpc, the AAU stream is three times further away, and therefore \gaia proper motion measurements are not available for stream members along the main sequence. Fortunately, the radial velocities provided by the spectroscopic measurements allow us to reliably remove the foreground contamination and present a clean sample of member stars in the streams, making it possible to detect extremely low surface brightness features created by perturbations.

With \SSSSS we have obtained spectroscopic data on over ten stellar streams (Paper I), some of which present relatively narrow stream widths, whose progenitors are likely to be globular clusters like the AAU stream.
The combination of photometric, astrometric and spectroscopic data will enable crucial new studies of the possible perturbation signatures in these streams.

\acknowledgments

This paper includes data obtained with the Anglo-Australian Telescope in Australia. We acknowledge the traditional owners of the land on which the AAT stands, the Gamilaraay people, and pay our respects to elders past and present.
This paper includes data gathered with the 6.5 meter Magellan Telescopes located at Las Campanas Observatory, Chile. 

We thank Paul McMillan for providing the posterior MCMC chains of the fit from \cite{McMillan:2017}.

TSL and APJ are supported by NASA through Hubble Fellowship grant HST-HF2-51439.001 and HST-HF2-51393.001, respectively, awarded by the Space Telescope Science Institute, which is operated by the Association of Universities for Research in Astronomy, Inc., for NASA, under contract NAS5-26555.  SK 
is partially supported by NSF grants AST-1813881, AST-1909584 and 
Heising-Simons foundation grant 2018-1030. 
ABP is supported by NSF grant AST-1813881.

This research has made use of the SIMBAD database, operated at CDS, Strasbourg, France \citep{Simbad}, and NASA's Astrophysics Data System Bibliographic Services.

This paper made use of the Whole Sky Database (wsdb) created by Sergey Koposov and maintained at the Institute of Astronomy, Cambridge by Sergey Koposov, Vasily Belokurov and Wyn Evans with financial support from the Science \& Technology Facilities Council (STFC) and the European Research Council (ERC).

This work presents results from the European Space Agency (ESA) space
mission Gaia. Gaia data are being processed by the Gaia Data
Processing and Analysis Consortium (DPAC). Funding for the DPAC is
provided by national institutions, in particular the institutions
participating in the Gaia MultiLateral Agreement (MLA). The Gaia
mission website is https://www.cosmos.esa.int/gaia. The Gaia archive
website is https://archives.esac.esa.int/gaia.

This project used public archival data from the Dark Energy Survey
(DES). Funding for the DES Projects has been provided by the
U.S. Department of Energy, the U.S. National Science Foundation, the
Ministry of Science and Education of Spain, the Science and Technology
Facilities Council of the United Kingdom, the Higher Education Funding
Council for England, the National Center for Supercomputing
Applications at the University of Illinois at Urbana-Champaign, the
Kavli Institute of Cosmological Physics at the University of Chicago,
the Center for Cosmology and Astro-Particle Physics at the Ohio State
University, the Mitchell Institute for Fundamental Physics and
Astronomy at Texas A\&M University, Financiadora de Estudos e
Projetos, Funda{\c c}{\~a}o Carlos Chagas Filho de Amparo {\`a}
Pesquisa do Estado do Rio de Janeiro, Conselho Nacional de
Desenvolvimento Cient{\'i}fico e Tecnol{\'o}gico and the
Minist{\'e}rio da Ci{\^e}ncia, Tecnologia e Inova{\c c}{\~a}o, the
Deutsche Forschungsgemeinschaft, and the Collaborating Institutions in
the Dark Energy Survey.  The Collaborating Institutions are Argonne
National Laboratory, the University of California at Santa Cruz, the
University of Cambridge, Centro de Investigaciones Energ{\'e}ticas,
Medioambientales y Tecnol{\'o}gicas-Madrid, the University of Chicago,
University College London, the DES-Brazil Consortium, the University
of Edinburgh, the Eidgen{\"o}ssische Technische Hochschule (ETH)
Z{\"u}rich, Fermi National Accelerator Laboratory, the University of
Illinois at Urbana-Champaign, the Institut de Ci{\`e}ncies de l'Espai
(IEEC/CSIC), the Institut de F{\'i}sica d'Altes Energies, Lawrence
Berkeley National Laboratory, the Ludwig-Maximilians Universit{\"a}t
M{\"u}nchen and the associated Excellence Cluster Universe, the
University of Michigan, the National Optical Astronomy Observatory,
the University of Nottingham, The Ohio State University, the OzDES
Membership Consortium, the University of Pennsylvania, the University
of Portsmouth, SLAC National Accelerator Laboratory, Stanford
University, the University of Sussex, and Texas A\&M University.
Based in part on observations at Cerro Tololo Inter-American
Observatory, National Optical Astronomy Observatory, which is operated
by the Association of Universities for Research in Astronomy (AURA)
under a cooperative agreement with the National Science Foundation.

The Pan-STARRS1 Surveys (PS1) and the PS1 public science archive have been made possible through contributions by the Institute for Astronomy, the University of Hawaii, the Pan-STARRS Project Office, the Max-Planck Society and its participating institutes, the Max Planck Institute for Astronomy, Heidelberg and the Max Planck Institute for Extraterrestrial Physics, Garching, The Johns Hopkins University, Durham University, the University of Edinburgh, the Queen's University Belfast, the Harvard-Smithsonian Center for Astrophysics, the Las Cumbres Observatory Global Telescope Network Incorporated, the National Central University of Taiwan, the Space Telescope Science Institute, the National Aeronautics and Space Administration under Grant No. NNX08AR22G issued through the Planetary Science Division of the NASA Science Mission Directorate, the National Science Foundation Grant No. AST-1238877, the University of Maryland, Eotvos Lorand University (ELTE), the Los Alamos National Laboratory, and the Gordon and Betty Moore Foundation.

TSL would like to thank Day for his company during the quarantine while accomplishing this manuscript.

{\it Facilities:} 
{Anglo-Australian Telescope (AAOmega+2dF); Magellan/Clay (MIKE)}

{\it Software:} 
{\code{numpy} \citep{numpy}, 
\code{scipy} \citep{scipy}, 
\code{matplotlib} \citep{matplotlib}, 
\code{astropy} \citep{astropy,astropy:2018}, \code{emcee} \citep{Foreman_Mackey:2013},  \code{CarPy} \citep{Kelson03}, 
\code{MOOG} \citep{Sneden73,Sobeck11}, \code{smhr} \citep{Casey14},
\code{q3c} \citep{Koposov2006}, 
\code{RVSpecFit} \citep{rvspecfit}
\code{GPyOpt} \citep{Gpyopt2016},
\code{STAN} \citep{Carpenter2017}
}


\bibliography{main}{}
\bibliographystyle{aasjournal}


\appendix

\section{Coordinate Transformation Matrix}\label{sec:coords} 

The transformation from celestial coordinates ($\alpha, \delta$) to the stream coordinates ($\phi_1, \phi_2$) is given by \citep{Shipp2019}:

\begin{eqnarray}
\begin{bmatrix}
\cos(\phi_1) \cos(\phi_2)\\
\sin(\phi_1) \cos(\phi_2)\\
\sin(\phi_2)
\end{bmatrix}&=&\nonumber\\
\begin{bmatrix}
0.83697865 & 0.29481904 & -0.4610298\\
0.51616778 & -0.70514011 & 0.4861566\\ 
0.18176238 & 0.64487142 & 0.74236331
\end{bmatrix} 
&\times &
\begin{bmatrix}
\cos(\alpha) \cos(\delta)\\
\sin(\alpha) \cos(\delta)\\
\sin(\delta)
\end{bmatrix}\nonumber
\end{eqnarray}

\section{Example of bar and spiral arm perturbations}\label{sec:spiral}

\begin{figure}
    \centering
    \includegraphics[width=0.98\textwidth]{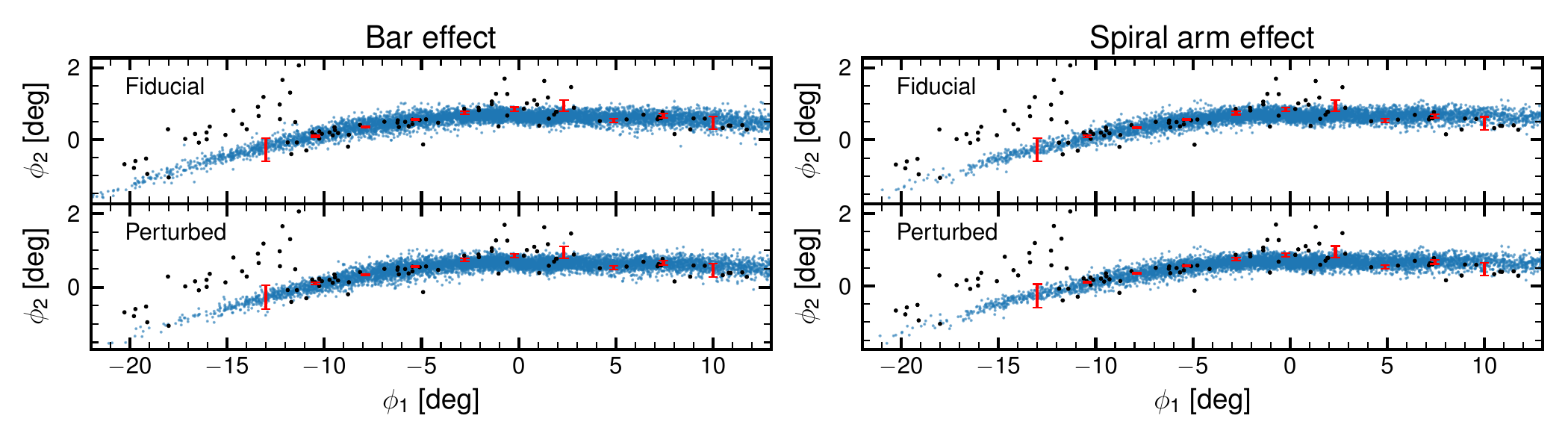}
    \caption{Example of perturbations from bar (left) and spiral arms (right) which produce the largest change in the stream track. \textit{Left panels} show the effect of the Milky Way bar. The top panel shows the fiducial bar simulation with a pattern speed of $\Omega = 1000\,\kms\,\mathrm{kpc}^{-1}$ and the bottom panel shows the perturbed stream evolved in the presence of a bar with pattern speed $\Omega = 42.3\,\kms\,\mathrm{kpc}^{-1}$. The maximum deviation is $0.1\degr$. \textit{Right panels} show the effect of spiral arms on ATLAS. The top panel shows the fiducial spiral arm simulation with a pattern speed of $\Omega = 1000\,\kms\,\mathrm{kpc}^{-1}$ while the bottom panel shows the simulation with the largest track deviation with a pattern speed of $\Omega = 26.6\,\kms\,\mathrm{kpc}^{-1}$. The largest deviation in the track is $0.02\degr$ showing that the spiral cannot create any appreciable features in ATLAS.}
    \label{fig:atlas_bar_spiral}
\end{figure}

In Section \ref{sec:bar}, \ref{sec:spiral_arms} we considered the effect of the Milky Way bar and spiral arm respectively. Both of these can create only modest perturbations in the stream. In Figure \ref{fig:atlas_bar_spiral} we show the stream realizations with the largest changes in the stream track ($0.1\degr$ for the bar and $0.02\degr$ for the spiral arms). 

\section{Example of globular cluster perturbations}\label{sec:gc}

\begin{figure}
    \centering
    \includegraphics[width=0.49\textwidth]{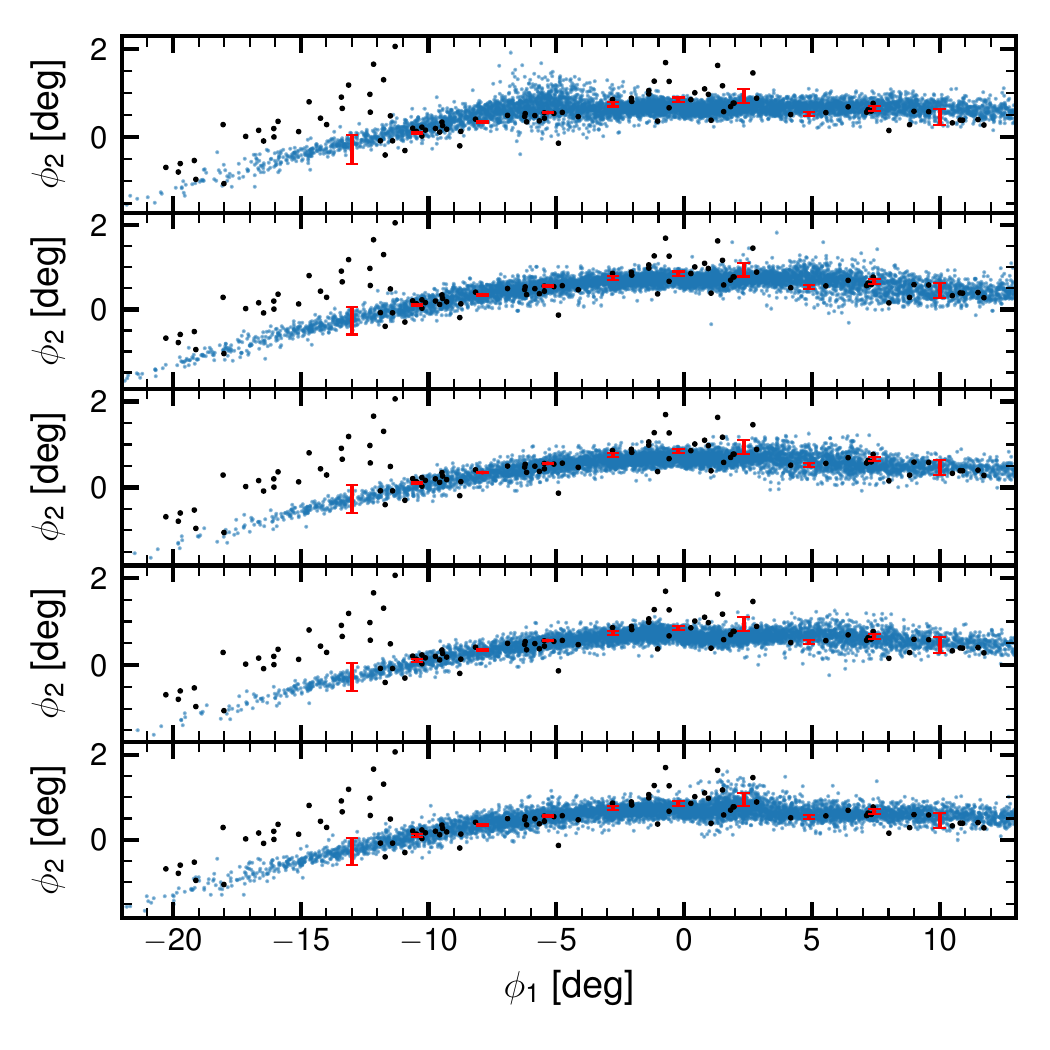}
    \caption{Example of perturbations from NGC 7492 to the AAU stream. We show five perturbations out of the 100 sampled in Section \protect\ref{sec:gc_perturbations}. These were chosen due to the change in the stream track as well as the broadening in the stream width. The top panel shows the perturbation with the largest change in the stream track, producing a kink with a size of 0.19$\degr$ at $\phi_1\sim-5\degr$ and an associated broadening of the stream. This is qualitatively similar to the wiggle and broadening observed at $\phi_1 \sim 3\degr$. }
    \label{fig:atlas_NGC7492}
\end{figure}

In Section \ref{sec:gc_perturbations} we explored the effect of globular clusters on the AAU stream. Of these, NGC 7492 has the closest approach to AAU with a median approach distance of 0.55 kpc. In Figure \ref{fig:atlas_NGC7492} we show five examples perturbations from NGC 7492. While none of these create kinks as large as the one between ATLAS and Aliqa Uma, several perturbations create smaller wiggles in the stream as well as broadening of the stream width qualitatively consistent with the observed wiggle and broadening at $\phi_1 \sim 3\degr$.

\end{document}

%% file: authors.tex

\author[0000-0002-9110-6163]{Ting~S.~Li}\email{tingli@carnegiescience.edu}
\altaffiliation{NHFP Einstein Fellow}
\affiliation{Observatories of the Carnegie Institution for Science, 813 Santa Barbara St., Pasadena, CA 91101, USA}
\affiliation{Department of Astrophysical Sciences, Princeton University, Princeton, NJ 08544, USA}
\author[0000-0003-2644-135X]{Sergey~E.~Koposov}
\affiliation{McWilliams Center for Cosmology, Carnegie Mellon University, 5000 Forbes Ave, Pittsburgh, PA 15213, USA}
\affiliation{Institute for Astronomy, University of Edinburgh, Royal Observatory, Blackford Hill, Edinburgh EH9 3HJ, UK}
\affiliation{Institute of Astronomy, University of Cambridge, Madingley Road, Cambridge CB3 0HA, UK}
\affiliation{Kavli Institute for Cosmology, University of Cambridge, Madingley Road, Cambridge CB3 0HA, UK}
\author[0000-0002-8448-5505]{Denis~Erkal}
\affiliation{Department of Physics, University of Surrey, Guildford GU2 7XH, UK}
\author[0000-0002-4863-8842]{Alexander~P.~Ji}
\altaffiliation{Hubble Fellow}
\affiliation{Observatories of the Carnegie Institution for Science, 813 Santa Barbara St., Pasadena, CA 91101, USA}
\author[0000-0003-2497-091X]{Nora~Shipp}
\affiliation{Department of Astronomy \& Astrophysics, University of Chicago, 5640 S Ellis Avenue, Chicago, IL 60637, USA}
\affiliation{Kavli Institute for Cosmological Physics, University of Chicago, Chicago, IL 60637, USA}
\affiliation{Fermi National Accelerator Laboratory, P.O.\ Box 500, Batavia, IL 60510, USA}
\author[0000-0002-6021-8760]{Andrew~B.~Pace}
\affiliation{McWilliams Center for Cosmology, Carnegie Mellon University, 5000 Forbes Ave, Pittsburgh, PA 15213, USA}
\author{Tariq~Hilmi}
\affiliation{Department of Physics, University of Surrey, Guildford GU2 7XH, UK}
\author[0000-0003-0120-0808]{Kyler~Kuehn}
\affiliation{Lowell Observatory, 1400 W Mars Hill Rd, Flagstaff,  AZ 86001, USA}
\affiliation{Australian Astronomical Optics, Faculty of Science and Engineering, Macquarie University, Macquarie Park, NSW 2113, Australia}
\author[0000-0003-3081-9319]{Geraint~F.~Lewis}
\affiliation{Sydney Institute for Astronomy, School of Physics, A28, The University of Sydney, NSW 2006, Australia}
\author[0000-0002-6529-8093]{Dougal~Mackey}
\affiliation{Research School of Astronomy and Astrophysics, Australian National University, Canberra, ACT 2611, Australia}
\author[0000-0002-8165-2507]{Jeffrey~D.~Simpson}
\affiliation{School of Physics, UNSW, Sydney, NSW 2052, Australia}
\author[0000-0002-3105-3821]{Zhen~Wan}
\affiliation{Sydney Institute for Astronomy, School of Physics, A28, The University of Sydney, NSW 2006, Australia}
\affiliation{Sydney Institute for Astronomy, School of Physics, A28, The University of Sydney, NSW 2006, Australia}
\author[0000-0003-1124-8477]{Daniel~B.~Zucker}
\affiliation{Department of Physics \& Astronomy, Macquarie University, Sydney, NSW 2109, Australia}
\affiliation{Macquarie University Research Centre for Astronomy, Astrophysics \& Astrophotonics, Sydney, NSW 2109, Australia}
\author[0000-0001-7516-4016]{Joss~Bland-Hawthorn}
\affiliation{Sydney Institute for Astronomy, School of Physics, A28, The University of Sydney, NSW 2006, Australia}
\affiliation{Centre of Excellence for All-Sky Astrophysics in Three Dimensions (ASTRO 3D), Australia}
\author[0000-0001-8536-0547]{Lara~R.~Cullinane}
\affiliation{Research School of Astronomy and Astrophysics, Australian National University, Canberra, ACT 2611, Australia}
\author{Gary~S.~Da~Costa}
\affiliation{Research School of Astronomy and Astrophysics, Australian National University, Canberra, ACT 2611, Australia}
\author{Alex~Drlica-Wagner}
\affiliation{Fermi National Accelerator Laboratory, P.O.\ Box 500, Batavia, IL 60510, USA}
\affiliation{Department of Astronomy \& Astrophysics, University of Chicago, 5640 S Ellis Avenue, Chicago, IL 60637, USA}
\affiliation{Kavli Institute for Cosmological Physics, University of Chicago, Chicago, IL 60637, USA}
\author[0000-0001-6924-8862]{Kohei~Hattori}
\affiliation{McWilliams Center for Cosmology, Carnegie Mellon University, 5000 Forbes Ave, Pittsburgh, PA 15213, USA}
\author[0000-0002-3430-4163]{Sarah~L.~Martell}
\affiliation{School of Physics, UNSW, Sydney, NSW 2052, Australia}
\affiliation{Centre of Excellence for All-Sky Astrophysics in Three Dimensions (ASTRO 3D), Australia}
\author[0000-0002-0920-809X]{Sanjib~Sharma}
\affiliation{Sydney Institute for Astronomy, School of Physics, A28, The University of Sydney, NSW 2006, Australia}
\affiliation{Centre of Excellence for All-Sky Astrophysics in Three Dimensions (ASTRO 3D), Australia}

\collaboration{20}{(\SSSSS Collaboration)}

%% file: table_specmem.tex
\begin{table*}
\begin{center}
\caption{
A total of 96 spectroscopic members in ATLAS stream and Aliqa Uma stream. Only first few lines are shown here. Full table is available in the online version in machine readable format.}
\scriptsize
\label{table:atlas_members}
\begin{tabular}{l r r r r r r r r}
\hline
\gaia DR2 Source ID  & RA   & Decl.  & SNR &  $G$ & $v_\mathrm{hel}$  &  $\sigma_v$\  & \feh  &  $\sigma_\feh$\ \\
 & (deg)  & (deg) &  & (mag)  & (\kms)  &  (\kms)  &    &    \\
\hline
\hline
2362404846580059648 &   9.109642 & -20.418631 &  21.2  & 16.48  &  -148.16 &  2.91 &       &       \\
2362395599515154816 &   9.387846 & -20.461972 &   3.7  & 19.56  &  -142.35 &  7.34 &       &       \\
2350314513642106624 &   9.890383 & -20.839192 &  17.9  & 17.72  &  -148.30 &  1.28 & -2.28 &  0.26 \\
2350310424833246592 &   9.974475 & -20.892894 &  32.5  & 16.83  &  -156.83 &  0.90 &       &       \\
2350245137034340864 &  10.193825 & -21.129194 &   7.2  & 18.26  &  -147.61 &  3.77 &       &       \\
2350348972163836160 &  11.151796 & -21.480250 &  36.5  & 16.55  &  -144.90 &  0.80 & -2.17 &  0.15 \\
2349548630777593344 &  11.609142 & -22.164725 &   8.8  & 18.50  &  -141.97 &  2.20 & -2.35 &  0.53 \\
2349572579516107264 &  11.654787 & -21.817247 &   4.9  & 18.99  &  -142.44 &  5.29 &       &       \\
2349268564550587904 &  12.229042 & -22.749461 &  22.9  & 16.38  &  -134.19 &  1.02 & -2.58 &  0.19 \\
\hline
\end{tabular}
\end{center}
\end{table*}

%% file: table_distance.tex
\begin{table}
\begin{center}
\caption{
BHB and RRL members in ATLAS stream and Aliqa Uma stream, together with the derived distance modulus. Only first few lines are shown here. Full table is available in the online version in machine readable format.}
\scriptsize
\label{table:distance}
\begin{tabular}{l r r r l}
\hline
\gaia DR2 Source ID  & RA   & Decl.  & $m-M$  & tracer \\
 & (deg)  & (deg) &   (mag)  &    \\
\hline
\hline
5033819563470215296	& 18.654975	& $-$26.549133	& 16.61	& rrl \\
5039633604864050048 & 19.678817	& $-$26.591158	& 16.71	& bhb \\
4969932298603707776	& 34.759075	& $-$33.963078	& 17.21	& bhb \\
4970235699391286016	& 33.274133	& $-$33.535181	& 17.16	& bhb \\
\hline
\end{tabular}
\end{center}
\end{table}

%% file: table_palca_specmem.tex
\begin{table*}
\begin{center}
\caption{
A total of 25 spectroscopic members in Palca stream, found in AAU stream fields. Only first few lines are shown here. Full table is available in the online version in machine readable format.}
\scriptsize
\label{table:palca_members}
\begin{tabular}{l r r r r r r r r}
\hline
\gaia DR2 Source ID  & RA   & Decl.  & SNR &  $G$ & $v_\mathrm{hel}$  &  $\sigma_v$\  & \feh  &  $\sigma_\feh$\ \\
 & (deg)  & (deg) &  & (mag)  & (\kms)  &  (\kms)  &    &    \\
\hline
\hline
4969841627550660608 &  33.905829 & -34.599403 &  18.9  & 17.74  &    90.77 &  1.47 & -1.94 &  0.19 \\
4969992221987294976 &  34.187021 & -33.840422 &   3.7  & 19.50  &   116.11 &  5.99 &       &       \\
4970601905482914048 &  34.795417 & -33.121203 &  36.8  & 16.54  &   117.48 &  0.79 & -2.06 &  0.13 \\
4970326404802896896 &  35.433371 & -33.664853 &   2.9  & 19.50  &    82.03 & 11.32 &       &       \\
4969997517683640832 &  34.555904 & -33.705767 &  75.3  & 15.81  &    89.33 &  0.69 & -1.82 &  0.11 \\
\hline
\end{tabular}
\end{center}
\end{table*}